\documentclass[onecolumn,english,aps,prb,twocolum,superscriptaddress,bibnotes,amsmath,amssymb,floatfix]{revtex4-2}
\usepackage[colorlinks=true,citecolor=blue,linkcolor=magenta]{hyperref}

\usepackage{graphicx}
\usepackage{epsfig}
\usepackage{color}
\usepackage{amssymb}
\usepackage{amsmath}
\usepackage{array}
\usepackage[loose]{units}
\usepackage{braket}
\usepackage{soul}
\usepackage{hyperref}
\usepackage{cleveref}
\usepackage[caption=false]{subfig}
\usepackage[english]{babel}
\usepackage{letltxmacro}
\usepackage{siunitx}
\usepackage{ dsfont }
\LetLtxMacro{\ORIGselectlanguage}{\selectlanguage}
\makeatletter

\newcommand{\strath}{SUPA and Department of Physics, University of Strathclyde, Glasgow G4 0NG, Scotland, UK}
\newcommand{\nice}{Universit\'{e} C\^{o}te d'Azur, CNRS, Institut de Physique de Nice, 06560 Valbonne, France}

\begin{document}

\title{Self-Organization in Cold Atoms Mediated by Diffractive Coupling}

\author{Thorsten Ackemann}
\email{thorsten.ackemann@strath.ac.uk}

\affiliation{\strath}
\author{Guillaume Labeyrie}
\affiliation{\nice}
\author{Giuseppe Baio}
\affiliation{\strath}
\author{Ivor~Kre\v{s}i\'{c}}
\thanks{Current address: Institute for Theoretical Physics, Vienna University of Technology, Vienna A-1040, Austria}

\affiliation{\strath}
\author{Josh G. M. Walker}
\affiliation{\strath}
\author{Adrian Costa Boquete}
\affiliation{\strath}
\author{Paul Griffin}
\affiliation{\strath}
\author{William J. Firth}
\affiliation{\strath}
\author{Robin Kaiser}
\affiliation{\nice}
\author{Gian-Luca Oppo}
\affiliation{\strath}
\author{Gordon R.M. Robb}
\affiliation{\strath}
\maketitle
\date{\today}




\textbf{This article discusses self-organization in cold atoms via light-mediated interactions induced by feedback from a single retro-reflecting mirror. Diffractive dephasing between the pump beam and the spontaneous sidebands selects the lattice period. Spontaneous breaking of the rotational and translational symmetry occur in the 2D plane transverse to the pump. We elucidate how diffractive ripples couple sites on the self-induced atomic lattice. The nonlinear phase shift of the atomic cloud imprinted onto the optical beam is the parameter determining coupling strength.  The interaction can be tailored to operate either on external degrees of freedom leading to atomic crystallization for thermal atoms and supersolids for a quantum degenerate gas, or on internal degrees of freedom like populations of the excited state or Zeeman sublevels. Using the light polarization degrees of freedom on the Poincar{\'e} sphere (helicity and polarization direction), specific irreducible tensor components of the atomic Zeeman states can be coupled leading to spontaneous magnetic ordering of states of dipolar and quadrupolar nature. The requirements for critical interaction strength are compared for the different situations. Connections and extensions to longitudinally pumped cavities, counterpropagating beam schemes and the CARL instability are discussed.}

\section{Introduction}
Spontaneous self-organization of near or out-of equilibrium states to ordered spatial structures has attracted considerable attention in many areas of science, technology and natural phenomena due to its appealing beauty, its importance in understanding how non-trivial structures can arise from homogeneous driving and the potential for applications \cite{turing52,prigogine67,busse78,meinhardt82,cross93}. In nonlinear optics, optical pattern formation or `transverse nonlinear optics' was investigated intensively in the 1980s-2000s, the observation by Grynberg of hexagons in counterpropagating beams traversing hot sodium vapour probably being the seminal experiment \cite{grynberg88b}. In addition to atomic vapours, liquid crystals, hybrid system such as liquid crystal light valves, photorefractives and semiconductors were used as nonlinear media. Configurations included longitudinally pumped cavities, single-mirror feedback schemes and counterpropagating beams. We refer to \cite{lugiato94b,vorontsov95,rosanov96,lange98a,arecchi99,ackemann01,staliunas03a,mandel04} for reviews. The common feature of these systems is that self-organization occurs in the plane transverse to the propagation direction of the pump beam and that the spatial coupling is mediated by diffraction.

A second strand in optical self-organization stems from the proposition of Rodolfo Bonifacio for collective atomic recoil lasing (CARL) \cite{bonifacio94a,bonifacio94}. This constitutes self-organization in the longitudinal direction (parallel to the pump propagation direction)  due to the bunching of atoms in optomechanical potentials, i.e.\ using external degrees of freedom of atoms compared to internal degrees of freedom which transverse pattern formation had usually focused on. Initial claims of observations in hot vapours remained controversial due to competing mechanisms providing similar signatures \cite{hemmer96,lippi96,verkerk97,lippi97,brown97a} but CARL was observed experimentally using cold atom clouds in single-mode ring cavities \cite{kruse03,voncube04,robb04} and has remained a very active area of research since then, as evidenced by this special issue.

A third strand in self-organization in atomic and optical systems arises from the considerable interest in achieving effective interaction between internal or external degrees of freedom of atoms via light modes to explore new self-organized quantum phases and the dynamics of complex quantum many-body systems by well-controllable model systems based on cold atoms. One workhorse has been the transversely pumped cavity, originally proposed in \cite{domokos02}. The self-organization discussed relies on external degrees of freedom as atoms bunch into a checkerboard lattice via an optical potential formed by the interference between the transverse pump and the spontaneously emerging cavity field. The transition has been demonstrated experimentally for laser-cooled thermal \cite{black03} and quantum degenerate atomic clouds \cite{baumann10}. In the latter case, a mapping to the quantum Dicke phase transition \cite{nagy10,baumann10} is possible and the resulting state can be interpreted as a (discrete) supersolid \cite{mottl12}. Reviews of the field can be found in Refs.~\cite{ritsch13,kirton19,mivehvara21}. If the cavity supports only a single spatial mode, the intra-cavity field provides a global coupling between atoms. In order to enable more local coupling there has been considerable interest in exploring multi-mode configurations either via degenerate cavities \cite{gopalakrishnan09,gopalakrishnan10,kollar17} or single-mode cavities with crossed axes \cite{leonard17}, the latter giving a continuous supersolid in quasi-1D. All these schemes have a limitation in the sense that there are two (or more) distinguished axes which imposes restrictions on the structures which can emerge from self-organization.

In contrast, self-organization in diffractively coupled systems with a single distinguished pump axis allows for spontaneous symmetry breaking of the rotational and translational symmetry in the plane transverse to the pump axis, at least for the idealized case of pumping with a plane wave. The use of cold atoms as a nonlinear medium was proposed for a counterpropagating beam scheme in \cite{muradyan05} and predicted the possibility of optomechanical bunching in the transverse plane due to dipole forces without the need for an intrinsic optical nonlinearity. First experimental clues of such an effect in a very low aspect ratio situation were obtained in \cite{greenberg11} and investigated further in \cite{greenberg12a,schmittberger14,schmittberger16a}. We are going to analyze a related system, originally proposed by Firth \cite{firth90a,dalessandro91}, in which the counterpropating beam is generated by a retro-reflecting mirror. This has the advantage that the length scale of the instability can be adapted by adjusting the distance between the atomic cloud and the feedback mirror.  The spontaneous formation of hexagonal atomic density patterns due to optomechanically mediated interaction was demonstrated in \cite{labeyrie14,tesio14}. Extensions to quantum degenerate gases were investigated theoretically in Refs.~\cite{robb15,zhang18} and might open up the possibility for the realization of a 2D supersolid. In addition to the novel feature of coupling to the centre-of-mass degrees of freedom, cold atomic gases are a highly controllable  medium with a high nonlinearity and negligible Doppler broadening offering the possibility to revisit self-organization based on internal degrees of freedom in a fruitful way. Structures due to the electronic 2-level nonlinearity were identified in \cite{camara15}. Spontaneous magnetic ordering of dipolar and quadrupolar nature was demonstrated in \cite{kresic18,kresic19} and \cite{labeyrie18}, respectively, the latter providing also a first demonstration of a spontaneously structured atomic coherence.

In this contribution, we will review the mechanism of diffractive coupling in single mirror feedback systems and provide a unified treatment of the different situations and the requirements on optical density and nonlinear phase shift. In particular, we will discuss how the light degrees of freedom on the Poincar{\'e} sphere (helicity and polarization direction) can couple to specific magnetic moments inside the atomic vapour and thus allow investigation of dipole and quadrupolar magnetic states. We will elucidate how diffractive ripples are coupling sites on the self-induced atomic lattice and discuss the long-range vs.\ short-range nature of the interaction. This supports an interpretation of atoms interacting with each other via light-mediated coupling. In contrast, the main emphasis in optical pattern formation until recently had been the coupling between optical waves mediated by the medium and often the media degrees of freedom were adiabatically eliminated formally. This approach persisted for the early studies of cold atomic media \cite{greenberg12a,schmittberger14,schmittberger16a}. After discussion of the single mirror feedback case, we provide a brief coverage of cold atom structures in longitudinally pumped cavities \cite{tesio12}. We close with a brief outlook and a discussion of the connection between different systems and the CARL instability.


\section{Mechanism of diffractive coupling}
\subsection{Single-mirror feedback schemes and the Talbot effect}\label{sec:principle}
\begin{figure}[h]
\includegraphics[width=15cm]{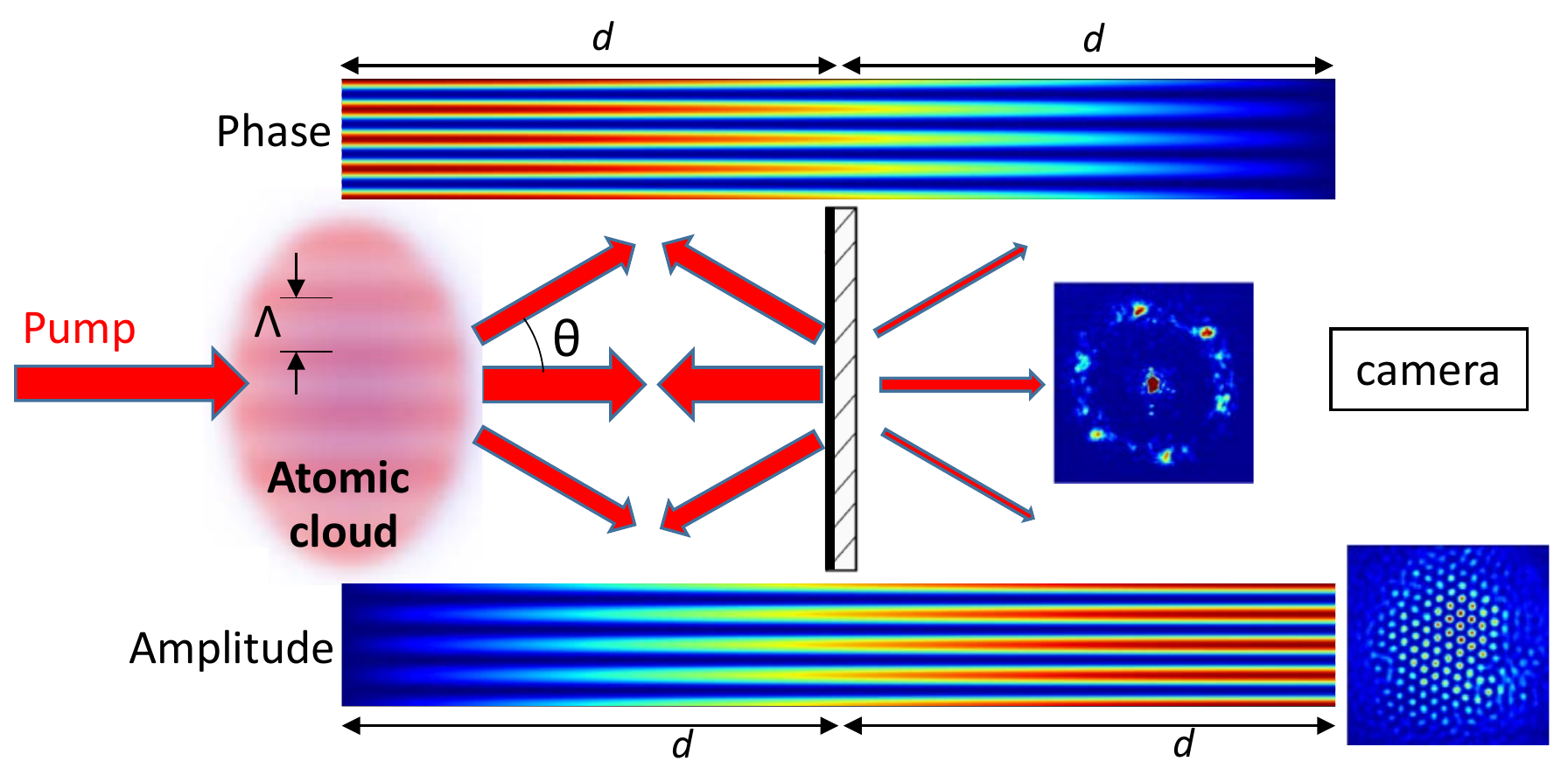}
\caption{ \label{fig:Talbot}
Scheme of feedback experiment: A coherent pump beam drives an atomic cloud of size $L$ with a plane feedback mirror at a distance $d$. Pump photons are scattered by a spatial structure with period $\Lambda$ in the cloud into sidebands at an angle $\theta$. Interference of these sidebands forms an optical lattice sustaining the spatial structure in the cloud. The sidebands can be visualized by looking at the far field of the transmitted light as obtained in the focal plane of a lens. The inset shows an example for an experimentally observed far field (FF) intensity distribution showing transverse Fourier space (displayed span -8.5 to 9.5 mrad around optical axis).  The upper panel illustrate how the phase modulation arising after the cloud is diminishing during propagation whereas an amplitude modulation (lower panel) builds up by diffractive dephasing.  At a distance $2d$ the amplitude modulation is fully developed and the phase is flat. Here, the transmitted light has the same spatial structure as the retro-reflected feedback beam re-entering the atomic cloud. This near field (NF) intensity distribution can be imaged onto a camera (example shown in lower right inset, displayed range 2 mm). Parameters for images: input intensity 129 mW/cm$^2$, detuning $\Delta = 7 \Gamma$, saturation parameter $s \approx 0.18$, dominantly optomechanical nonlinearity.
}
\end{figure}

\begin{figure}[h]
\centering
\includegraphics[width=7.5cm]{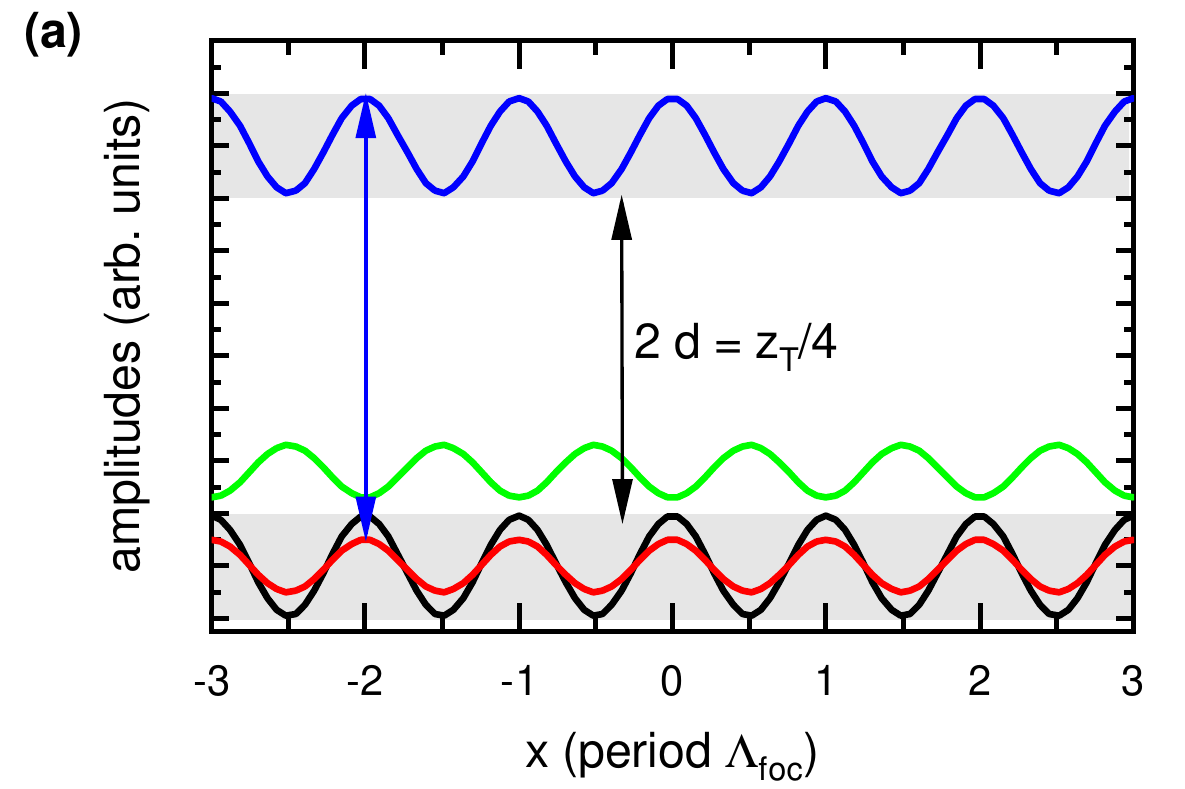} \includegraphics[width=7.5cm]{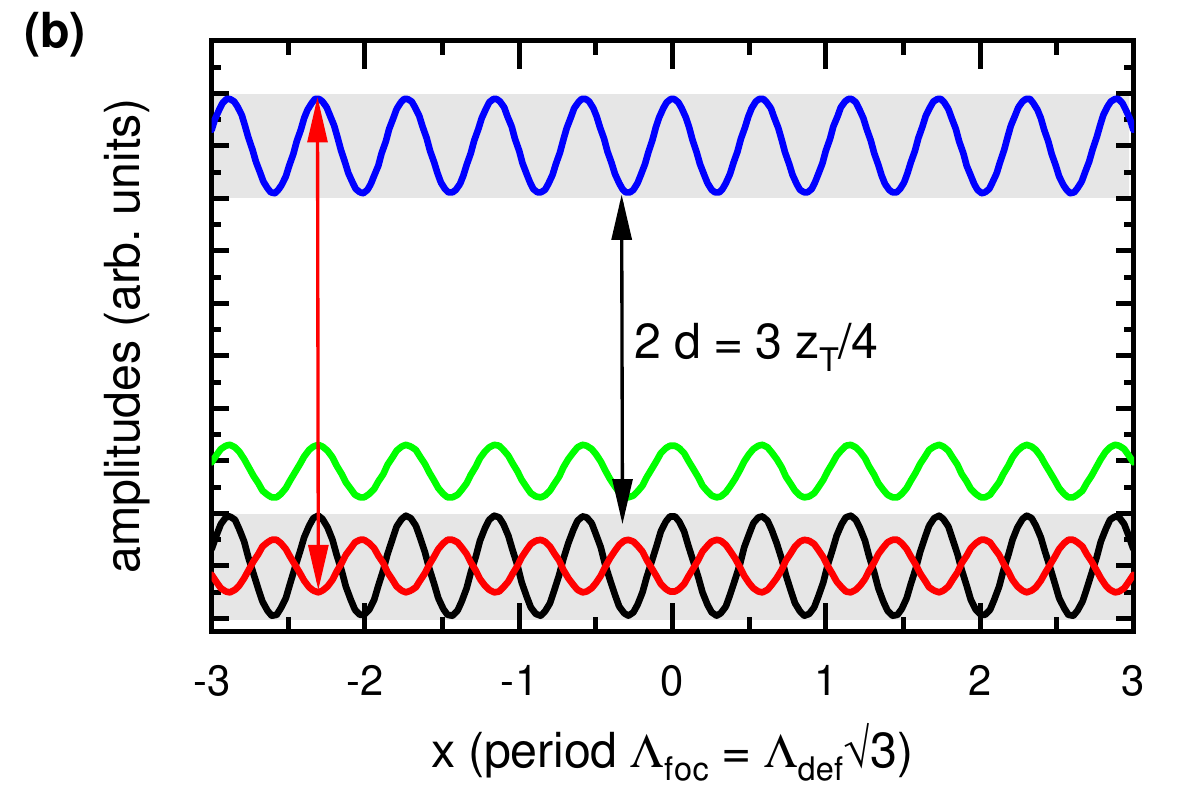}
\caption{\label{fig:feedback_electronic} Illustration of feedback loop for single mirror feedback. The $x$-axis shows transverse space in periods of the pattern. The $y$-axis denotes amplitude, but via the vertical displacements also propagation distance.  The two light grey regions represent the atomic cloud as encountered by the forward beam (bottom) and the backwards beam reentering the cloud (top) after propagating to the mirror and back (distance $2d$, indicated by black arrow).  The thick black line indicates a modulation of an atomic state variable, here for concreteness the population $\rho_{ee}$ of the excited state. The red line shows the resulting modulation of the refractive index. The green line indicates the phase of the transmitted beam just after traversing the cloud. The blue line indicates the amplitude modulation created in the  backward beam by diffractive dephasing.
\textbf{(a)} Blue detuning: After a quarter of the Talbot distance intensity maxima in the backward beam are aligned with refractive index maxima (blue arrow) providing positive feedback for a self-focusing nonlinearity.
\textbf{(b)} Red detuning: After a three quarters of the Talbot distance intensity maxima in the backward beam are aligned with refractive index minima (red arrow) providing positive feedback for a self-defocusing nonlinearity. Hence the period of the spatial modulation in the cloud experiencing maximum positive feedback for a given mirror distance $d$ is reduced by a factor $\sqrt{3}$.
}
\end{figure}

A powerful scheme for transverse optical pattern formation is the single mirror feedback arrangement \cite{firth90a,dalessandro91} illustrated in Fig.~\ref{fig:Talbot}. A cloud of atoms of size $L$ is driven by a coherent broad-area beam (a plane wave in the theoretical conception). The transmitted light is retro-reflected by a plane feedback mirror placed at a distance $d$ behind the centre of the medium. The common assumption, which we make also here, is that the cloud is considerably thicker than the light wavelength ($L\gg \lambda$) but diffractively thin, i.e.\ $L \ll d$. These assumptions lead to a conceptually simple situation. Within the medium only the nonlinearity is present but no diffraction, whereas in the feedback loop the light experiences  diffraction only. From the experimental point of view the system is attractive as it does not demand interferometric stability of the mirror position as for a cavity. Any fluctuation of a state variable (density, excited state population, or population or coherences in the ground state Zeeman levels, black line in Fig.~\ref{fig:feedback_electronic}) will change the refractive index of the cloud (red line in Fig.~\ref{fig:feedback_electronic}) and thus the transmitted light is phase modulated (green line in Fig.~\ref{fig:feedback_electronic}). Diffraction in the feedback loop is governed by the paraxial wave equation expressed in real space or Fourier space as:
\begin{eqnarray}
    \frac{\partial}{\partial z} E(x,y,z) &=&-\frac{i}{2k} \Delta_\perp E(x,y,z), \label{eq:parax_real}\\
  \tilde{E}(q_x,q_y,z+\delta z) &=& exp{\left(\frac{iq^2}{2k}\,\delta z\right)}\,  \tilde{E}(q_x,q_y,z), \label{eq:parax_Fourier},
\end{eqnarray}
where $E(x,y,z)$ denotes the slowly varying optical field, $\Delta_\perp$ the transverse Laplace operator, $\tilde{E}(q_x,q_y,z)$ its transverse Fourier transform, $q=\sqrt{q_x^2+q_y^2}$ the transverse wavenumber and $k=2\pi /\lambda$ the total wavenumber. The diffractive phasor in (\ref{eq:parax_Fourier}) describes the dephasing between the pump and any spontaneous spatial sideband created and leads to a periodic conversion between phase and amplitude modulation. It is at the origin of the Talbot effect stating that every periodic field modulation with wavenumber $q$ reproduces itself at the Talbot distance $z_T=4\pi k/q^2$ \cite{talbot36}. A pure phase modulation after the cloud will be converted to a pure amplitude modulation (to first order) after a quarter of the Talbot distance (\cite{ciaramella93}, see the illustrations on top and bottom of Fig.~\ref{fig:Talbot} and Fig.~\ref{fig:feedback_electronic}) and, directed back into the cloud by the mirror, can then reinforce the original perturbation providing positive feedback for an instability growing spontaneously from quantum or thermal noise. At a quarter of the Talbot distance positive feedback is provided for a self-focusing situation, i.e.\ one in which the refractive index increases with intensity (blue line in Fig.~\ref{fig:feedback_electronic}\textbf{(a)}). The spatial period selected out of the noise spectrum is
\begin{equation}\label{eq:foc}
  \Lambda_{foc}= \sqrt{4\lambda d} .
\end{equation}
For a self-defocusing medium, i.e.\ one in which the refractive index decreases with increasing intensity, the field needs to propagate to three quarters of the Talbot distance for positive feedback (blue line in Fig.~\ref{fig:feedback_electronic}\textbf{(b)}) as the diffractive phasor changes sign then. For a given mirror distance, this implies a smaller critical period,
\begin{equation}\label{eq:def}
  \Lambda_{def}= \sqrt{\frac{4}{3}\lambda d} .
\end{equation}

\begin{figure}[h]
\includegraphics[width=15cm]{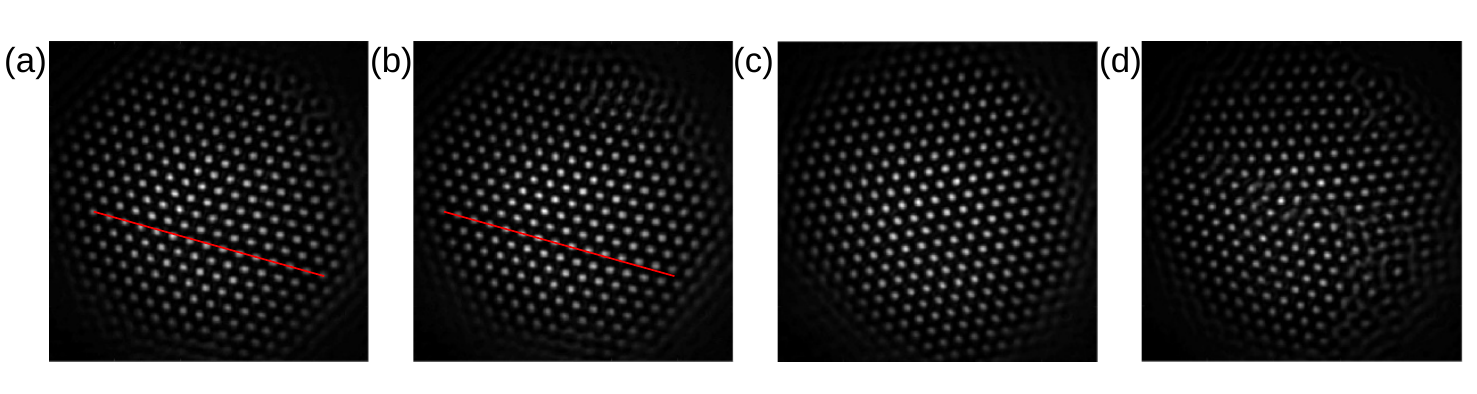}
\caption{\label{fig:symbreaking}Examples of hexagonal patterns observed in a parameter range in which optomechanical and electronic nonlinearities are both present. The different structures appeared in different realizations of the experiment under nominally the same experimental conditions illustrating spontaneous symmetry breaking. \textbf{(a)}, \textbf{(b)}  and \textbf{(c)} differ in the orientation of hexagonal axes. This difference is very small between \textbf{(a)} and \textbf{(b)} (the red line indicating an axis in \textbf{(a)} is not aligned with a row of spots in \textbf{(b)}), but very obvious between \textbf{(a)}/\textbf{(b)} and \textbf{(c)}. \textbf{(d)} contains two different domains consisting of the hexagons in \textbf{(a)} and \textbf{(c)} with a defect line in between them. Parameters: $I_\textrm{in} = 250\,$mW/cm$^2$, $\delta = +7\,\Gamma$, and $d=5$~mm.}
\end{figure}

The angle $\theta$ between the spontaneous sidebands and the pump (Fig.~\ref{fig:Talbot}) is given by
\begin{equation}\label{eq:angle}
  \theta =\frac{q}{k}=\frac{\lambda}{\Lambda},
\end{equation}
and is best visualized by monitoring the far field of the transmitted light (see lower inset on right hand side of Fig.~\ref{fig:Talbot}). The argument via the Talbot effect gives only the wavenumber and thus indicates only a ring in Fourier space. This can be seen as a faint background in between the hexagonal peaks in the centre inset of \ref{fig:Talbot}. However, spontaneous symmetry breaking selects only a few modes on this ring and in most situations hexagons are selected (we will discuss the reason for this in Sec.~\ref{sec:dipole}). The interference of these six sidebands leads then to a hexagonal structure in the cloud. This can be visualized by imaging the field intensity after the cloud on a camera (see lower inset on rhs of Fig.~\ref{fig:Talbot} and Fig.~\ref{fig:symbreaking}). Typical angles and length scales are $\theta \approx 6$~mrad and $\Lambda \approx 120$~$\mu$m, but can be varied between 100 and 250~$\mu$m by moving the mirror in a range of $|d|\approx 0\ldots 40$~mm \cite{labeyrie14}. It should be noted that the mirror is not placed directly at these distances as that would interfere with the vacuum chamber, but relayed by an afocal telescope as suggested first in \cite{ciaramella93} and termed `virtual mirror'. This implies that one can place the mirror `into' the cloud or even before it. In the latter case the length scale changes from defocusing to focusing or vice versa due to the additional sign change of modulation. We find that the structures can be qualitatively described by the `diffractively thin medium' theory discussed here, even if the medium is moderately `thick' ($L \approx 10$~mm for the setup operated at INPHYNI, $L \approx 3$~mm for the setup operated at Strathclyde). A quantitative description of length scales and thresholds demands a theory taking diffraction within the medium into account \cite{firth17}. The main result of this theory is that the period of the structures does not become arbitrarily small when $d$ is reduced but saturates roughly at $\sqrt{\lambda L}$ corresponding to about 90~$\mu$m for the Nice setup and 50~$\mu$m for the Strathclyde setup. This is also the length obtained in the situation where the cloud is driven by two independent counterpropagating beams \cite{firth90}. In a first approximation, the latter instability can be understood as comparable to a system of two thin slices separated by $L$, and hence to a single slice with a feedback mirror at distance $L/2$.

The emerging hexagons have the freedom to choose their orientation on the critical ring in Fourier space. Examples are shown in Fig.~\ref{fig:symbreaking}. Panels \textbf{(a)}, \textbf{(b)}  and \textbf{(c)} differ in the orientation of hexagonal axes. Panel \textbf{(d)} contains two different domains consisting of the hexagons in \textbf{(a)} and \textbf{(c)} with a defect line in between them. Here, fluctuations triggered different directions of symmetry breaking in different areas which could not reconcile until a macroscopic modulation was reached. The resulting defect line is metastable. We are not aware of quantitative investigations for hexagons but for stripe-like patterns evidence for the Kibble-Zurek mechanism \cite{kibble76,zurek85} for defect formation during spontaneous symmetry breaking was given in \cite{ducci99,labeyrie16}. The rotational symmetry is preserved in a Gaussian input beam, admitting for unavoidable experimental imperfections, but the translational symmetry present in a plane wave is broken by the Gaussian envelope and hence the position of the patterns is typically pinned, although small variations can be detected in large enough beams. A quantitative analysis of symmetry breaking of hexagons with weak confinement and constraints is a subject of interesting further work and could be undertaken with a spatial light modulator in the input beam path allowing control of imperfections.

\subsection{2-level systems: Kerr and saturable nonlinearities}
First experimental investigations in single-mirror systems were performed on liquid crystals \cite{macdonald92a,ciaramella93} but soon after a wealth of novel phenomena was obtained in atomic systems \cite{grynberg94,grynberg94a,ackemann94,ackemann95b}. These realizations used hot rubidium without a buffer gas \cite{grynberg94,grynberg94a} and sodium vapour in a buffer gas atmosphere \cite{ackemann94,ackemann95b}. In the former case the nonlinearity is high but Doppler broadening, the vicinity of multiple hyperfine lines and the ballistic motion of the atoms make detailed modelling very challenging. In the latter case, the motion of the atoms is diffusive and can be easily modelled by a partial differential equation. The pressure broadening was larger than the hyperfine splitting, but the light-matter coupling is significantly reduced by the collisional broadening. Both schemes relied on an optical pumping nonlinearity using the $D_1$-lines (described qualitatively as a $J=1/2\to J'=1/2$-transition) creating orientations in the ground state (see Sec.~\ref{sec:Zeeman}). Cold atoms are attractive as they allow interaction with (dominantly) only one hyperfine transition, and have small residual atomic motion and hence negligible Doppler broadening.  The cold atom experiments discussed here (and all others discussed in the following sections) was performed on the $F=2 \to F'=3$ hyperfine transition of the D$_2$-line of $^{87}$Rb \cite{labeyrie14,camara15}. The atomic cloud is prepared in a magneto-optical trap which is then switched off for the experiment. The temperature $T$ of the atoms is around the Doppler temperature, $T_{dop} \approx$ 150~$\mu$K, the exact value depending on the experimental conditions, in particular the size of the atomic cloud. For the clouds with very high optical density studied in this section and in Sec.~\ref{sec:optomech}, it is about 290~$\mu$K.  As saturation of a 2-level transition, i.e.\ of the electronic transition between the ground and excited state is the archetypical atomic nonlinearity, we are going to look into this first. In the rate equation approximation, the equation of motion for the population of the excited state $\rho_{ee}$ is
\begin{equation}
\label{eq:s} \dot{\rho}_{ee} = -\Gamma \rho_{ee} + P  -2 P\rho_{ee} ,
\end{equation}
where $\Gamma$ denotes the spontaneous radiative decay rate, and $P$ the pump rate proportional to intensity. The first term describes spontaneous decay, the second the pumping and the third saturation. The back action of the cloud on the transmitted field is given by
\begin{equation}
\frac{\partial}{\partial z} E(x,y,z) =\frac{i \chi_{lin}}{2k} (1-2\rho_{ee}) E(x,y,z) = \frac{i \chi_{lin}}{2k} \, \frac{1}{1+s}\, E(x,y,z),
\end{equation}
where
\begin{equation}\label{eq:chi_electron}
\chi_{lin} = \frac{b_0}{kL}\, \frac{\frac{2\Delta}{\Gamma} +i}{1+\left(\frac{2\Delta}{\Gamma}\right)^2}
\end{equation}
denotes the linear susceptibility, $b_0$ the optical density at line centre, $\Delta$ the detuning and $s=2P/\Gamma$ the saturation parameter. Due to the counterpropagation, there is a longitudinal grating at wavelength scale formed in the cloud. We assume  that the residual atomic motion, although small, is sufficient to wash out the effects of this grating and to add the forward and backward intensity to calculate $P$. With this assumption, a good qualitative and even semi-quantitative agreement between experiment and theory has been obtained for optomechanical and optical pumping nonlinearities \cite{labeyrie14,kresic18,kresic19}. (For the electronic nonlinearity, this assumption is questionable and the grating has been included in the treatment in \cite{firth17}, but it is instructive and qualitatively correct to develop the theory without the grating.) Then the homogeneous steady state of the system is given by
\begin{equation} \label{eq:hom}
\rho_{ee,h}= \frac{ P_0(1+Rf_h)}{\Gamma +2P_0(1+Rf_h)},\end{equation}
where $R$ denotes the mirror reflectivity, $P_0$ the input pump rate of the forward beam alone and \newline$f_h=\exp{(b_0(1-\rho_{ee,h})/(1+(2\Delta / \Gamma)^2)}$
the transmission function. A linear stability analysis against a spatially periodic perturbation $\exp{(iqx + \eta t)}$ yields a growth rate of \cite{ackemann06u}
\begin{equation} \label{eq:eta}
 \eta = -\Gamma -2 P_0(1+R f_h) + R f_h P_0 (1-2\rho_{ee,h})  \, \frac{2b_0}{1+\left(\frac{2\Delta}{\Gamma}\right)^2} \,\left(\frac{2\Delta}{\Gamma}  \sin{\Theta} +\cos{\Theta}\right) ,
 \end{equation}
where $\Theta= q^2 d/k$ denotes the diffractive phasor. The first term denotes damping by spontaneous emission, the second damping by power broadening. The third term is the driving. It is proportional to the mirror reflectivity $R$, the optical density $b_0$ and the forward pump rate $P_0$. It is reduced by absorption, as $Rf_h$ represents an effective reflectivity. Hence single-mirror feedback schemes are sensitive to absorption and need to be operated off-resonantly. (In principle absorptive pattern formation via the cosine term is possible \cite{glueckstad95,aumann99}, but for an electronic two-level nonlinearity threshold cannot be reached.)  For large enough detuning, $\Delta \gg \Gamma$, the wavenumber selection is dominantly given by the sine term, $\Theta=\pi/2$ for $\Delta >0$ (self-focusing) and $\Theta=3\pi/2$ for $\Delta <0$ (self-defocusing). As a final important observation, the driving term is also diminished by the homogeneous solution, i.e.\ the saturation. Under the assumption that $\Delta \gg \Gamma$ so that the absorptive terms can be neglected, a simplified version of the growth rate can be obtained
\begin{equation} \label{eq:eta_disp}
\eta = -\Gamma -2 P_0(1+R) + 2 R P_0 (1-2\rho_{ee,h})  \, \phi_{lin} ,
\end{equation}
where $\phi_{lin}= b_0\, \Delta /\Gamma\, /(1+(2\Delta /\Gamma )^2)$ is the linear phase shift by the cloud, the dispersive optical density, which determines the interaction strength.  The nonlinear phase shift induced on the forward beam by the action on the medium of forward and backward beams together is given by $\phi_{nonlin}= \phi_{lin} 2\rho_{ee}= \phi_{lin} s/(1+s)$
 (note $s$ denotes the saturation parameter due to both beams).
For a simple analytic threshold condition, it is necessary to neglect saturation, i.e.\ to apply the Kerr approximation used originally in  \cite{firth90a,dalessandro91}, resulting in $\phi_{nonlin}= \phi_{lin}s = \phi_{lin} 2P_{0}/\Gamma$. The Kerr threshold is given by $P_{0,th}/\Gamma =1/(2R\phi_{lin})$ and $s_{th}=1/(R\phi_{lin})$. Hence, the Kerr threshold is reached for a nonlinear phase shift of $\phi_{nonlin,th} = 1/R$, which approaches  1~rad for $R\to 1$.

\begin{figure}[h]
\includegraphics[width=8cm]{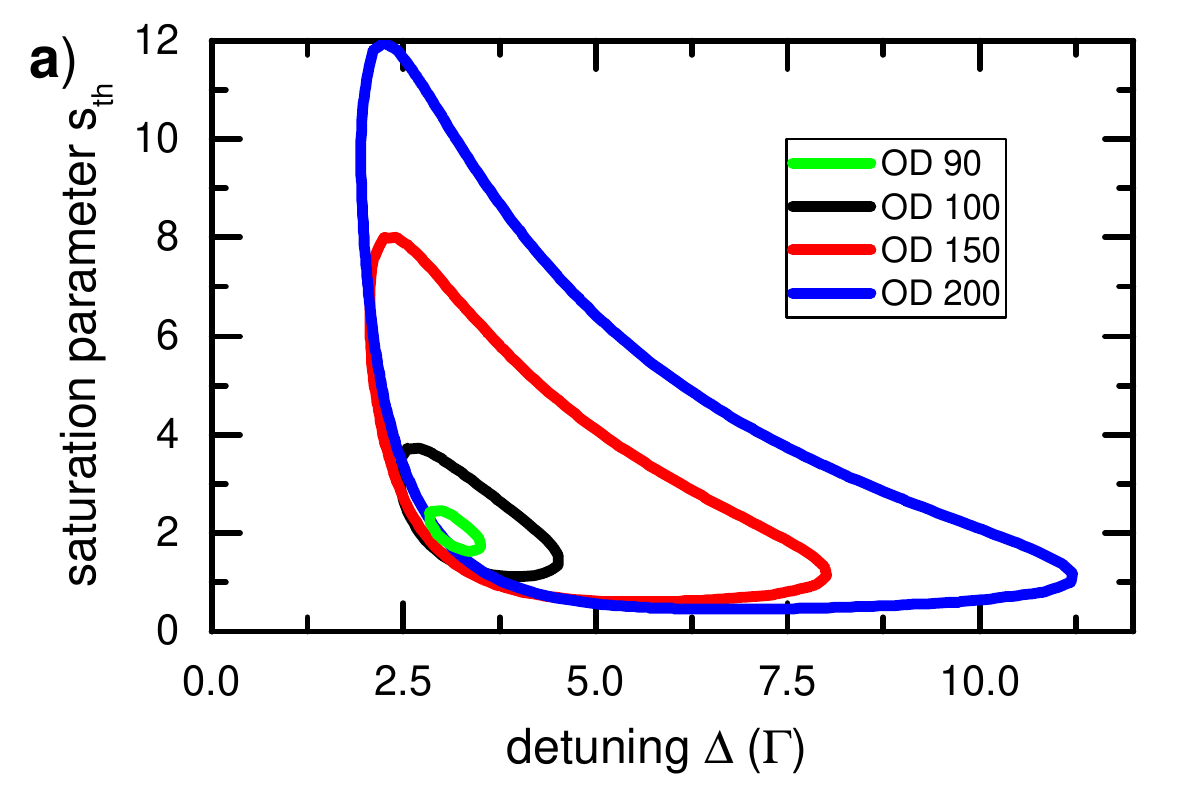}\includegraphics[width=8cm]{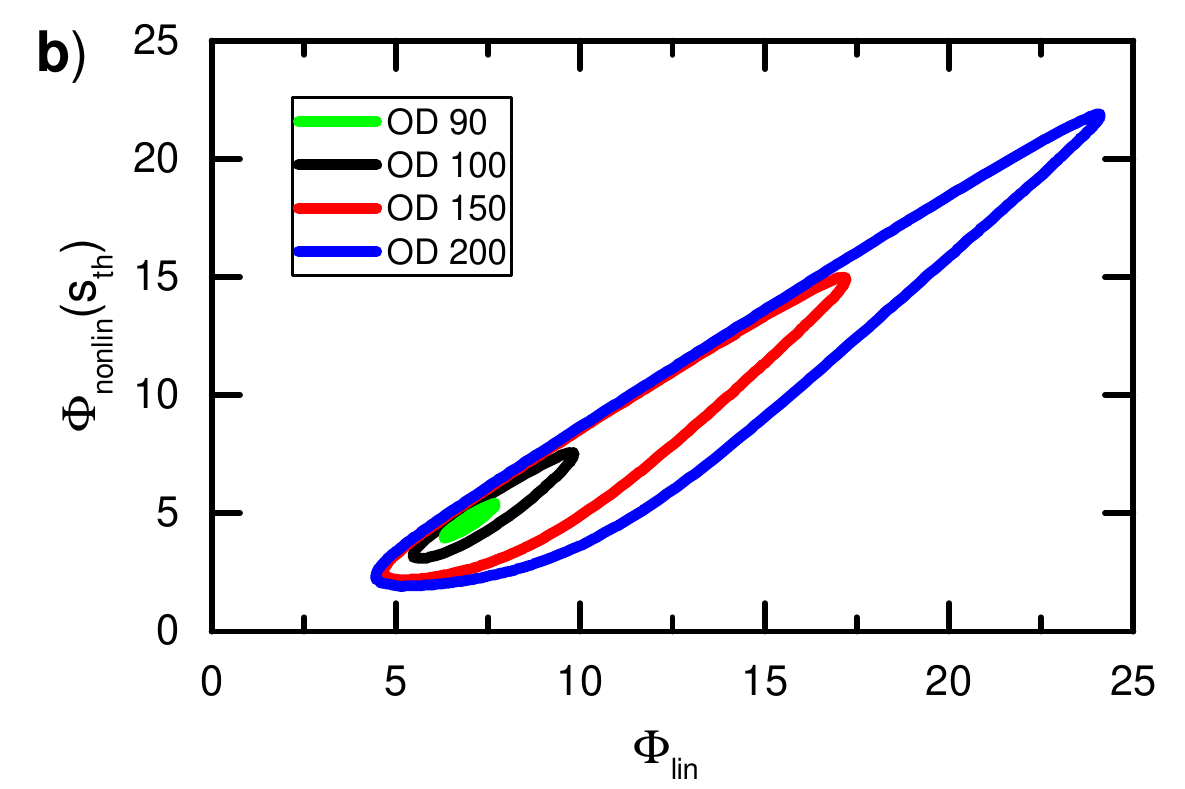}
\caption{\label{fig:2level_threshold}\textbf{(a)} Saturation parameter at threshold vs.\ detuning for four optical densities obtained from (\ref{eq:eta}) and (\ref{eq:hom}). \textbf{(b)} Nonlinear phase shift at threshold vs.\ linear phase shift (in radians). $s=P_0(1+f_h)$ denotes the saturation parameter including absorption. $R=1$ for simplicity. }
\end{figure}

\begin{figure}[h]
\centering
\includegraphics[width=9cm]{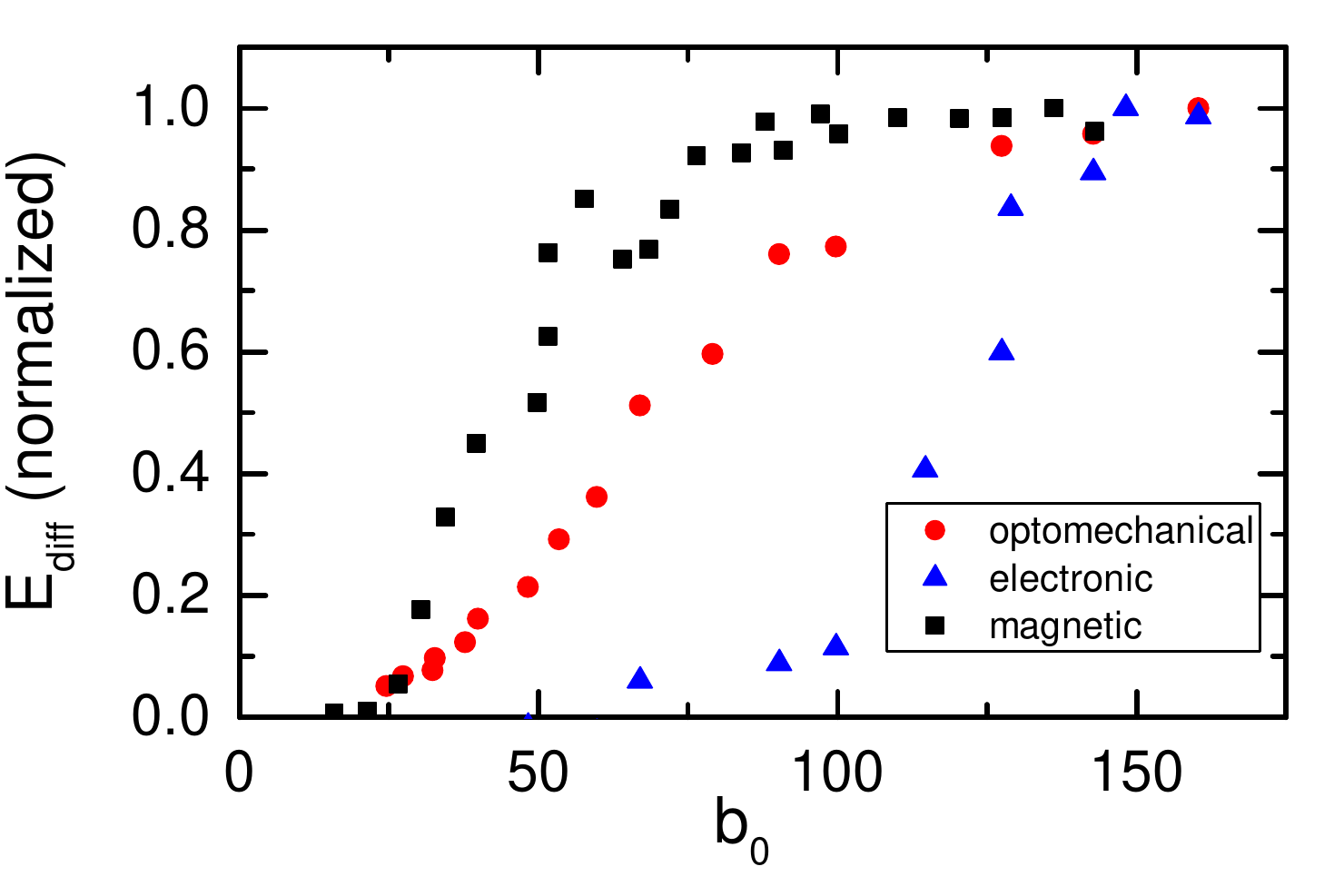}
\caption{\label{fig:OD_threshold}Diffracted energy in sidebands vs. optical density in line centre for optomechanical (red circles), electronic 2-level (blue triangles) and optical pumping nonlinearities (black squares). Each subset of data is normalized to its maximal value. For the optomechanical and electronic case the raw data are from Fig.~S1 of \cite{labeyrie14} and pulse duration is used to distinguished between them (see Sec.~\ref{sec:optomech}): short pulses ($1\,\mu$s, blue triangles),  long pulses ($200\,\mu$s, red circles). Parameters: $\Delta=+6\Gamma$, $I = 487\,$mW/cm$^2$. Optical density is varied by varying duration of thermal expansion of the cloud by introducing a controlled delay between  MOT turn-off  and pump pulse. For the optical pumping nonlinearity, parameters are $\Delta = 12 \Gamma$, $I = 3.6$~mW/cm$^2$, pump duration 0.5 ms.
The optical density is varied by adjusting the repumper power before the MOT is turned off.
}
\end{figure}

These features can be found in Fig.~\ref{fig:2level_threshold}\textbf{(a)}, which displays the threshold saturation parameter vs.\ detuning for different optical densities using the full theory including absorption and saturation. Threshold is minimal for an intermediate detuning as the phase shift diminishes for large detuning and absorption limits the range for small detunings, well before the maximum phase possible is reached at $\Delta=\Gamma/2$. The cut-off due to absorption at low detuning is pretty steep indicating the need for a high enough detuning in the experiment. Avoiding absorption also facilitates the theoretical treatment. There is not only a lower but also an upper threshold. This is due to the reduction of driving due to saturation of the homogeneous state in Eqs.~(\ref{eq:eta}), (\ref{eq:eta_disp}).  The instability is typically strongest for intermediate saturation. For low saturation, in the Kerr limit, the total nonlinear phase shift is not sufficient to reach threshold, whereas at high saturation, the differential positive feedback is not sufficient. The instability range increases with increasing optical density. The minimal optical density necessary for an electronic instability is about 90. This is the reason why it took relatively long to observe this instability in cold atoms \cite{labeyrie14,camara15} as achieving this density is not trivial for a spherical magneto-optical trap. (Cylindrical traps as used in \cite{greenberg11} will prevent large aspect ratio patterns.) We did not do systematic studies on the dependence of threshold on optical density, but a  threshold of an optical density of 90 is in good agreement with experiments as shown in Fig.~\ref{fig:OD_threshold}. It indicates that at constant input intensity, the diffracted energy into the sideband decays strongly below an optical density of 100.

Fig.~\ref{fig:2level_threshold}\textbf{(b)} presents the data of Fig.~\ref{fig:2level_threshold}\textbf{(a)} in a slightly different way. We plot the nonlinear phase shift by the combined action of forward and backward beam vs.\ the linear phase shift, i.e.\ the dispersive optical density. One can see that a nonlinear phase shift of about 2 rad is necessary to reach threshold, about double the value for a Kerr medium without saturation. This illustrates the detrimental effect of saturation. The minimal linear phase shift necessary is about 4.5 rad. This implies the requirement that $b_0 > 18 \Delta /\Gamma$ to reach threshold at a given detuning (large enough to avoid significant absorption). This is in qualitative agreement with the experiment and more detailed treatments in \cite{labeyrie14,camara15}.
For larger linear phase shift, the instability sets in at larger nonlinear phase shifts as absorption needs to be bleached.

It should be noted that these requirements for the optical density cannot be reduced by using a transition with a larger dipole matrix element as relaxation and driving increase both with increasing light-matter coupling. This will be different for the optomechanical and optical pumping nonlinearities discussed in the following sections. These considerations, in particular the ones on the limitations by saturation, are important also for liquid crystal-based systems, if saturation needs to be taken into account \cite{neubecker95}, or any other saturable atomic nonlinearity. A detailed analysis was performed for the optical pumping nonlinearity in a $J=1/2 \to J'=1/2$-hot vapour system in \cite{ackemann01b}. All the considerations discussed here are of relevance also for hot vapour systems as long as the assumption of homogeneous broadening can be justified, i.e.\ either by a sufficient pressure broadening to mask the Doppler broadening \cite{ackemann95b,ackemann01b}, or by a sufficiently large detuning, if no buffer gas is used. In the latter case the ballistic motion of the atoms might be modelled in a first approximation by a rate constant as in \cite{grynberg94,leberre95b}, but in general the ballistic motion will induce nonlocal effects \cite{skupin07}. In the former case, atomic motion can be modelled easily by introducing a diffusion constant \cite{ackemann95b,ackemann01b}. Note that the effect of diffusion can be quite strong. For example, the homogeneous relaxation constant for the orientation due to optical pumping in \cite{ackemann95b,ackemann01b,ackemann96t} was only about 10 s$^{-1}$ and atomic diffusion closed completely the instability region discussed for a saturable medium here. Self-organization was only possible by introducing artificial effective losses via a transverse magnetic field. We will discuss further aspects of the optical pumping nonlinearity in Sec.~\ref{sec:Zeeman}. We expect that the considerations developed here will provide useful guidance for planning cold as well as hot atom experiments.

\section{External degrees of freedom: Optomechanics} \label{sec:optomech}
\begin{figure}[h]
\includegraphics[width=7.5cm]{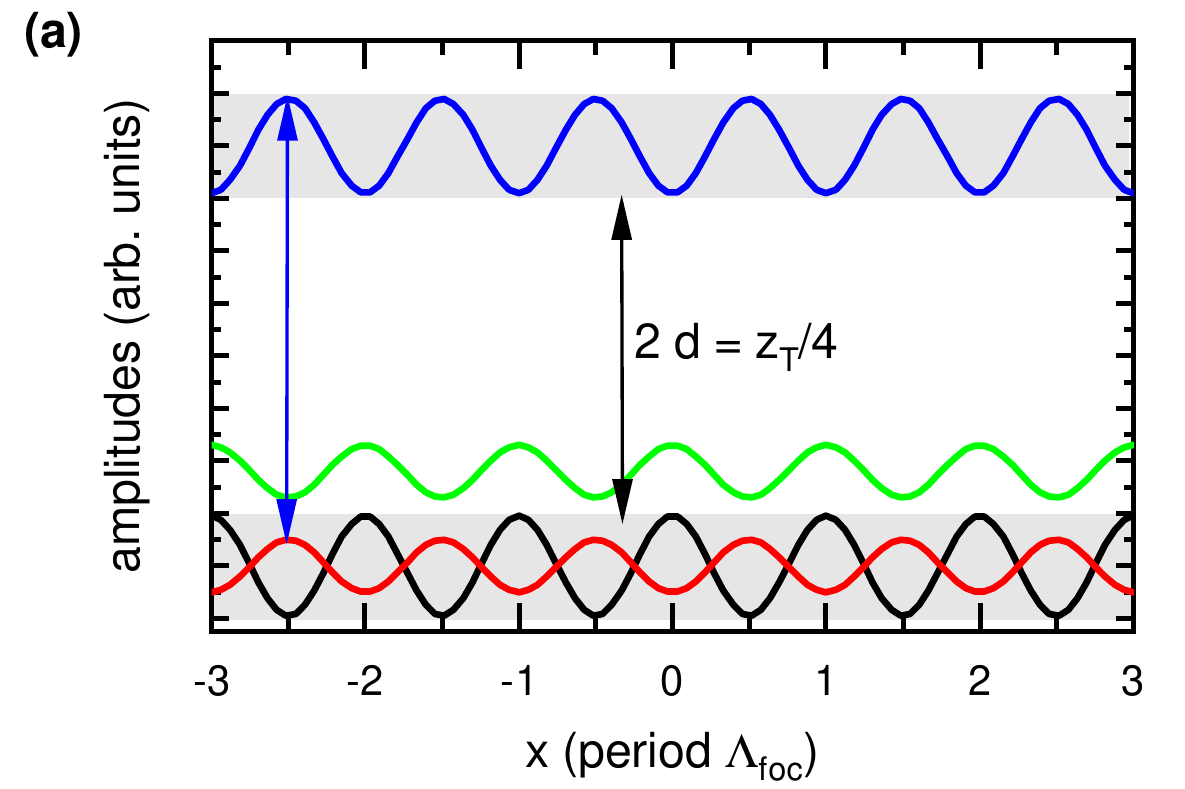}
\includegraphics[width=7.5cm]{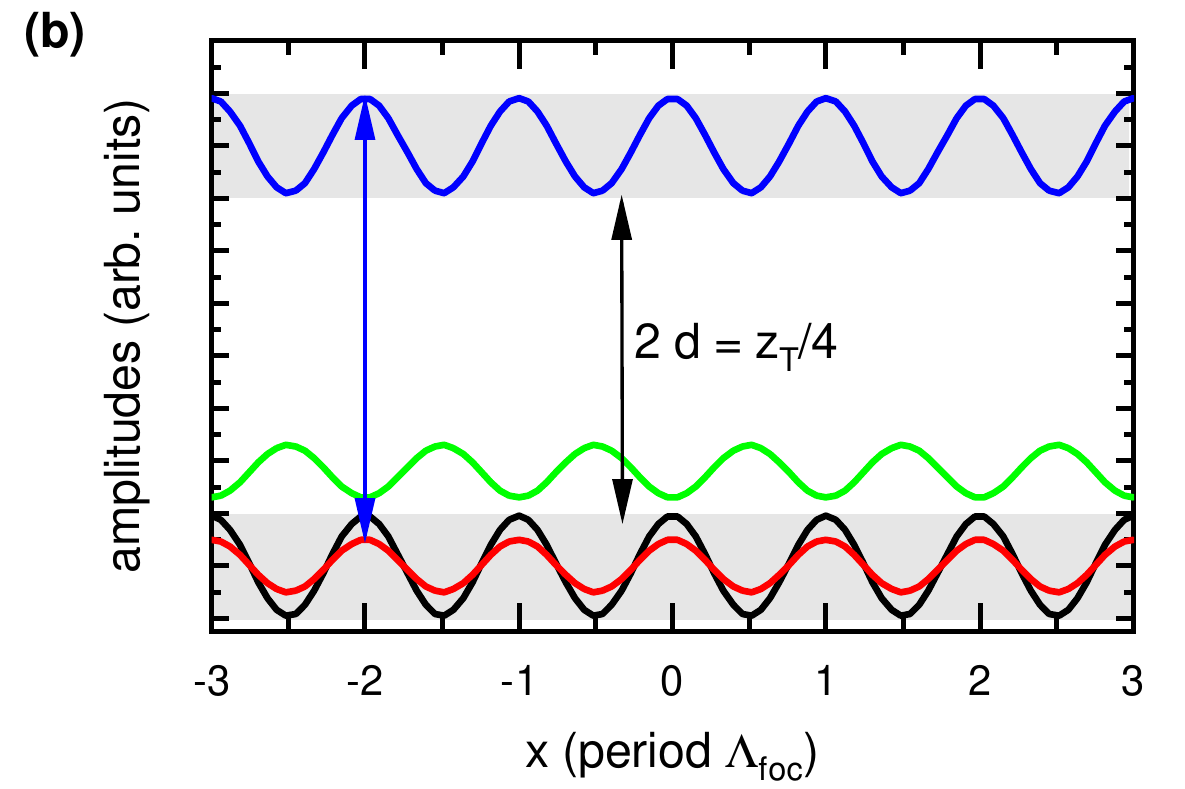}
\caption{\label{fig:feedback_optomech} Illustration of feedback loop for single mirror feedback for optomechanical interactions (see caption of Fig.~\ref{fig:feedback_electronic} for general explanation).  Black line in atomic cloud denotes atomic density.
\textbf{(a)} Blue detuning ($\Delta >0$): Atoms are low field seekers and hence atomic and intensity structures are complementary (atomic troughs align with intensity peaks, situation of experiment described in \cite{labeyrie14}).
\textbf{(a)} Red detuning ($\Delta <0$): Atoms are high field seekers and hence atomic and intensity structures are aligned.
Note that the nonlinearity is self-focusing and the instability exists for both signs of the detuning (similar to \cite{saffman98} for the case of single-pass propagation).}
\end{figure}

The qualitative new feature for cold atoms compared to hot atomic vapours is that the thermal velocity is so low, about 0.2 m/s at the Doppler temperature, that the motion of atoms is strongly affected by optical dipole forces.  The feedback mechanism is illustrated in Fig.~\ref{fig:feedback_optomech}. The black line denotes a modulation in the atomic density. For blue detuning (Fig.~\ref{fig:feedback_optomech}\textbf{(a)}), the refractive index of an atom is smaller than 1. Hence the refractive index increases where there are less atoms (red line). After a quarter of a Talbot distance, intensity maxima are aligned with density minima. As atoms under blue detuned excitation are low field seekers, this provides positive feedback. The nonlinearity is effectively self-focusing but one expects complementary patterns in the optical intensity and the atomic density. For red detuning (Fig.~\ref{fig:feedback_optomech}\textbf{(b)}), the refractive index increases with atomic density (red line). After a quarter of a Talbot distance, intensity maxima are aligned with density maxima. As atoms under red detuned excitation are high field seekers, this provides positive feedback. This is again a self-focusing situation and corresponds to the pioneering identification of a synthetic optical Kerr nonlinearity in dielectric beads due to optomechanical interaction by Ashkin \cite{ashkin82}. The mechanism is also analogous to the bunching obtained in the CARL instability and the transversely pumped cavities. Pump photons are scattered at the density structure within the medium into sidebands. The interference of these sidebands with the pump creates a self-induced optical lattice in which the atoms localize.

In the experiment in \cite{labeyrie14} blue detuning is utilized. The optical intensity is monitored in transmission via a camera yielding the striking hexagonal patterns depicted in the inset of Fig.~\ref{fig:Talbot} and in Fig.~\ref{fig:symbreaking}. The atomic structure is probed via dispersive imaging with a red detuned probe beam and yields a honeycomb structure, i.e.\ the atoms assemble at ridges of low intensity in between the hexagonal peaks, as expected from the argument given above. Importantly, this probe structure has a slow decay (about 80 $\mu$s) after the pump is switched off. This  indicates that it cannot stem from a population modulation in the excited state, as this should decay within a few times $1/\Gamma = 26$~ns.

The optomechanical and the electronic cases can be also distinguished by looking at time scales of the turn-on dynamics and threshold power. For intensities above about 400~mW/cm$^2$ ($\Delta =6-7 \Gamma$, $b_0=150$), the first images acquired a microsecond after the switch-on of the pump pulse contain already a structure in the optical intensity but not in the probe beam. Afterwards, contrast increases in both of them on time scales of tens of microsecond. For intensities below about 200~mW/cm$^2$, the intensity structures and probe structures emerge together. Minimum threshold obtained is about 50~mW/cm$^2$, i.e.\ the optomechanical threshold is about a factor of 5-6 lower than the electronic one.

The theoretical treatment in \cite{labeyrie14} is based on a Vlasov-type equation for the phase space distribution function $f=f(\vec{x}, \vec{v},t)$ (with $\vec{x}$,$\vec{v}$ being the position and velocity vectors in the transverse plane) driven by the dipole force
\begin{equation}
\frac{\partial f}{\partial t} + \vec{v}\cdot \frac{\partial f}{\partial \vec{x}} + \frac{\hbar \Delta}{2M}\, \frac{\partial \log(1+s(\vec{x},t))}{\partial \vec{x}} \cdot \frac{\partial f}{\partial \vec{v}} \, = \, 0,
\end{equation}
where $M$ denotes atomic mass. The spatial atomic density $\rho(\vec{x},t)$ is recovered by integrating $f$ over the velocity space. The back action of the cloud on the field is given now by the susceptibility
\begin{equation}\label{eq:chi_optomech}
\chi(\vec{x},t) = \frac{b_0}{kL} \, \frac{\frac{2\Delta}{\Gamma} +i}{1+\left(\frac{2\Delta}{\Gamma}\right)^2} \, \frac{\rho(\vec{x},t)}{1+s(\vec{x},t)},
\end{equation}
where $\rho$ is normalized so that it is 1 in the homogeneous state. The results are in good agreement with experiment. In particular the model confirms that the lower threshold is essentially determined by the optomechanical effect. However, the inclusion of absorption and electronic saturation prevents an analytic treatment. Hence, in \cite{tesio14} the purely optomechanical dispersive limit was analyzed. A analytical expression for the input saturation pump parameter is obtained
\begin{equation}
s_{th} = \left[\frac{\hbar\Delta}{k_B T}\, R\phi_{lin} -(1+R)\right] ^{-1} ,
\end{equation}
where $k_B$ denotes the Boltzmann constant. From this it is apparent that the threshold is determined by the residual kinetic energy of the atoms and can be arbitrarily small for low temperatures in the plane wave limit. Furthermore, the threshold saturation parameter is independent of detuning for large enough detuning (as $\phi_{lin}\sim 1/\Delta$). The threshold intensity increases quadratically with detuning. This is the same scaling as in transversely pumped cavities and in general expected for laser induced dipole-dipole interactions  \cite{donner19p,odell03}. For the parameters of \cite{labeyrie14}, $b_0=150$, $\Delta =7\Gamma$, $T=290$~$\mu$K, one obtains $\phi_{lin}=5.3$, $s_{th}\approx 0.03$, $I_{th}\approx 20$~mW/cm$^2$. The minimum linear phase shift needed at this temperature is $\phi_{lin}\approx 0.3$ and the minimum optical density in line centre is $b_0\approx 8$. These values are significantly lower than the requirements for the electronic instability. Indeed structures were found down to a value of  $b_0\approx 25$ (see Fig.~\ref{fig:OD_threshold}). The discrepancy of about a factor of 2-3 between minimum threshold intensity and minimum optical density in experiment and analytical prediction is attributed to residual absorption and heating in the blue detuned molasses present due to the counterpropagating beams.

As indicated already, Ashkin identified a synthetic Kerr nonlinearity for dielectric beads \cite{ashkin82} and it has been used previously to obtain solitons in single-pass propagation  \cite{smith81,man13} and 1D pattern formation in counterpropagating beams \cite{reece07}. Recently, an observation of 2D patterns in the feedback scheme discussed here was reported in \cite{bobkova21}. Compared to these soft matter systems, cold atoms have the advantage that the dynamics can be studied without viscous damping of motion, allowing for a dissipation free evolution with a Hamiltonian description of the system, as discussed above. As a result, the dynamics has a tendency to be oscillatory as atoms cannot lose their energy in the optical potential. However, there are mechanisms for nonlinear damping of the motion \cite{tesio14a}. If desired, it is possible to introduce dissipation in a controlled way via optical molasses as done for the CARL instability \cite{voncube04} and suggested for counterpropagating beams in \cite{muradyan05} and cavities in \cite{tesio12}. The threshold of this model is essentially identical to the conservative one at the same initial temperature \cite{tesio14t} but the long term dynamics are different, forming a true `crystal', and experimentally one can expect a much longer lifetime of the emerging structures. First experimental indications of the effect  were obtained \cite{gomes15u} but no systematic investigations were performed.

As the interaction via the dipole force is coherent, it should be possible to use a quantum degenerate gas as the medium. This question was addressed in \cite{robb15} by coupling a Bose-Einstein condensate (BEC) described by a Gross–Pitaevskii equation to a feedback mirror. In a first step to investigate the effect of the light-mediated coupling on the BEC, the scattering length $a$ is assumed to be zero, i.e.\ atomic interaction takes place via the external feedback only. The first observation is that the threshold for the quantum degenerate gas does not go to zero as the temperature goes to zero, but remains finite due to the zero point kinetic energy. The same conclusion had been obtained before for the CARL instability \cite{bonifacio05}. The emerging state combines spatial symmetry breaking to a periodic structure with the breaking of the SU(1) symmetry related to the superfluidity of the BEC and thus should have supersolid behaviour. Supersolidity has been elusive in the condensed matter He systems where it was originally proposed \cite{andreev69,kim04a,kim12} but attracted  a lot of attention recently due to the observation in dipolar gases \cite{kaudau16,boettcher19,chomaz19}, spin-orbit coupled BEC lattices \cite{li17} and crossed cavities \cite{leonard17}, all confined to quasi 1D geometries. Numerical simulations of the feedback system yield fairly regular modulations in 1D \cite{robb15}. 2D structures show partial hexagonal order and remain dynamic during time evolution \cite{robb19u}. This is probably related to the different nature of the instability. For the single-mirror feedback system with thermal atoms there is an instability at finite wavenumber while the zero wavenumber remains stable \cite{tesio14t,firth90a}. For the BEC case, the range of unstable wavenumbers extends to zero, but zero wavenumber is marginal and the maximum growth rate occurs at finite wavenumber (Fig.~2 of \cite{robb15}) similar to the modulational instability in the Nonlinear Schr\"odinger and Gross–Pitaevskii equation \cite{bespalov66,nguyen17}. For this situation, the resulting structures are known to be quite disordered in 2D (see, e.g., filamentation in nonlinear beam propagation \cite{abbi71}).

In \cite{robb15}, it is assumed that the scattering length $a$ is zero. An attractive interaction in the BEC will enhance the instability and a repulsive interaction will counteract it. It will be very interesting to study the interaction between the two instabilities as the feedback modes are scanned across the intrinsic modulational instability spectrum for $a<0$ by changing the mirror distance. It should be also noted that a mapping of the instability on the Dicke model is possible \cite{tesio14t} and the transition has been interpreted as a quantum phase transition in \cite{nagy10,baumann10}.

Analysis in \cite{zhang19} of a limiting case for $a>0$ indicates that a single droplet is the ground state of the system with regions of more complex arrangement of droplet clusters existing at the transition between homogenous and structured states. If allowing the forward and backward beam to interact with transitions in a three-level V-systems at different detunings, a wealth of further clustered and extended states including quasi-periodicity are predicted \cite{zhang20}. Pursuing novel quantum phases using the flexible and peculiar nature of the coupling in the single-mirror scheme (see Sec.~\ref{sec:coupling}) promises to be an agenda for further exciting research.


\section{Internal degrees of freedom: Magnetic ordering}\label{sec:Zeeman}

\subsection{Optical pumping nonlinearity and irreducible tensor components}
A small but important detail not mentioned in the discussion of the experiments \cite{labeyrie14,camara15} in the sections before is that a linear polarizer with a transmission axis parallel to the linear input polarization was present in the feedback loop. This inhibited polarization instabilities and provided the possibility of a quasi-scalar description of the instability. If this restriction is released, a new wealth of phenomena becomes accessible based on optical pumping nonlinearities \cite{kastler57}, i.e.\ nonlinearities relying on non-equilibrium populations and coherences in the Zeeman substates of the ground states. These have very long lifetimes, limited only by atomic collisions (very rare for cold thermal atoms), depolarization at the cell walls (also negligible for cold atoms) and stray magnetic fields. Hence phenomena will appear at even lower thresholds and lower optical densities, reducing the demands on the driving laser and MOT apparatus.

\begin{figure}[h]
\centering
\includegraphics[width=11cm]{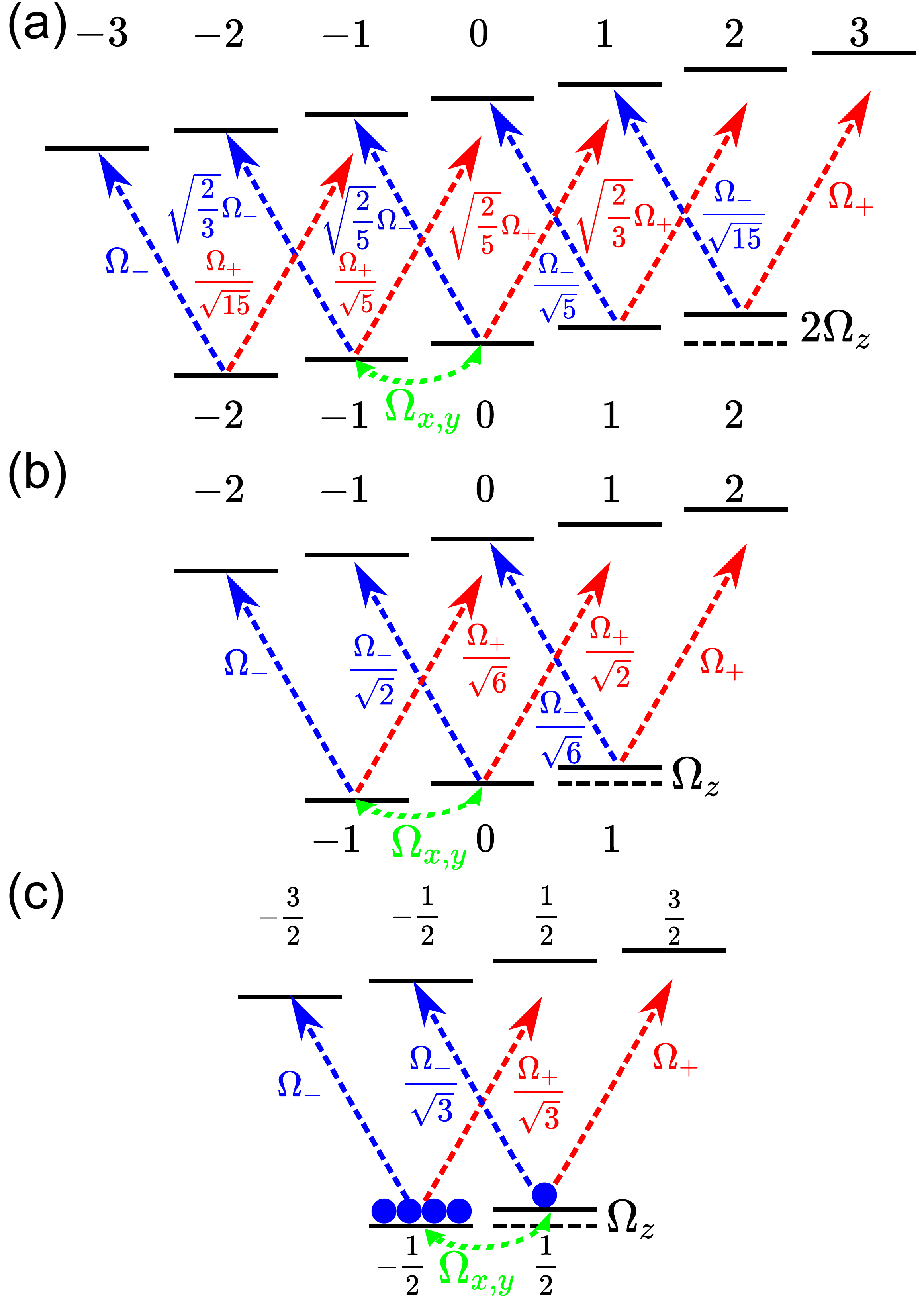}
\caption{\label{fig:levelscheme}\textbf{(a)} Level structure of $F=2\to F'=3$ hyperfine transition line of  $D_2$-line of $^{87}$Rb. Red (blue) arrows indicate transitions via $\sigma_+$ ($\sigma_-$) light with Rabi frequencies $\Omega_+$ ($\Omega_-$) with strength relative to the strongest transitions between the stretched states. A longitudinal magnetic field with Larmor frequency $\Omega_z$ splits the Zeeman sublevels. For simplicity, the $g-$factor for the ground state is assumed to apply also for the excited state. Transverse magnetic fields $\Omega_x$, $\Omega_y$ will induce coherences between Zeeman sublevels (for simplicity only one is shown).  \textbf{(b)} Reduced scheme of a $F=1\to F'=2$ transition used in the theoretical description. It includes the possibility of orientation (irreducible tensor rank 1) and alignment (rank 2) states in the ground state.  \textbf{(c)} Further reduced scheme of a $F=1/2\to F'=3/2$ transition keeping only the possibility of an orientation. In this situation, it is illustrated that an orientation with an increased population in the state with $m<0$ leads to a back-action on the light increasing absorption and phase shift for the $\sigma_-$-component. This orientation can be created by a dominance of $\sigma_-$- over $\sigma_+$-light, or for linearly polarized light by the incoherent Faraday effect. }
\end{figure}

The experiment is performed on the $F=2\to F'=3$ hyperfine transition line of  $D_2$-line of $^{87}$Rb, which has the Zeeman level structure depicted in Fig.~\ref{fig:levelscheme}\textbf{(a)}. $\sigma_+$ light will pump atoms towards magnetic substates with positive magnetic quantum numbers ending finally in the $m_F=2\to m_F'=3$ stretched state transition. This creates a positive orientation or magnetic dipole in the ground state. $\sigma_-$ light will have the opposite effect.

\begin{figure}[h]
\includegraphics[width=15cm]{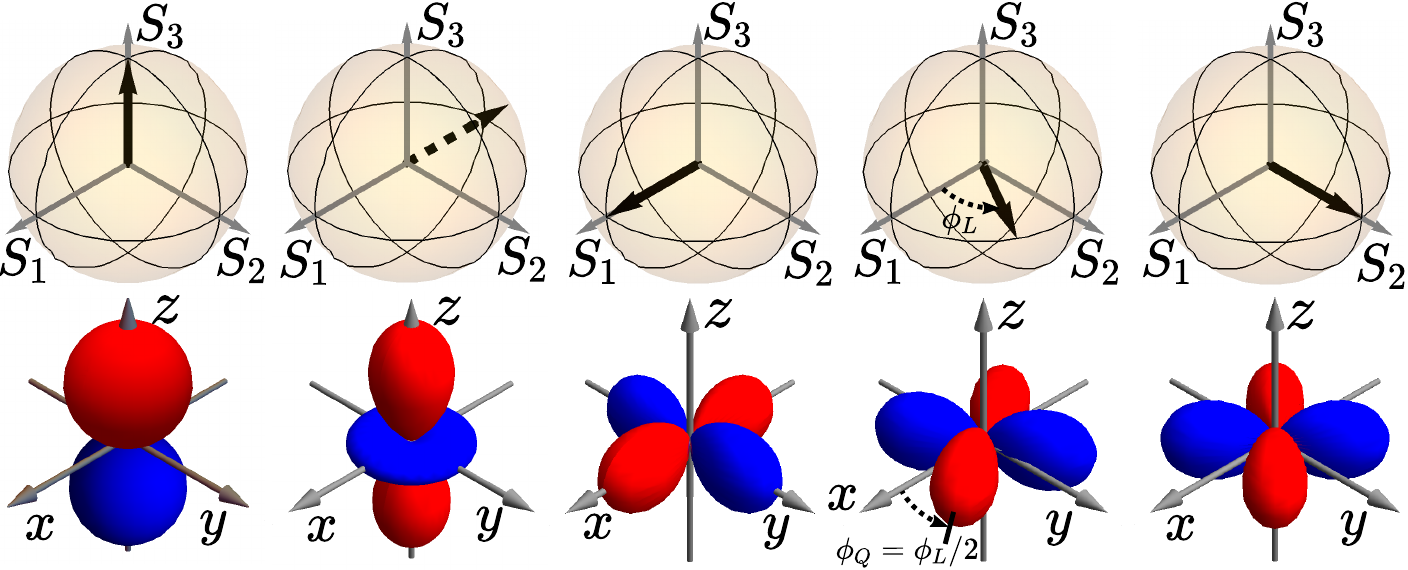}
\caption{\label{fig:moments} Polarization states on Poincar{\'e} sphere (upper row) and illustrations of magnetic states (lower row) coupling to them. Red denotes south pole, blue north pole (for positive irreducible components). Leftmost column: A light beam with  helicity, i.e.\ non-zero $S_3$ (out of the equatorial plane), will generate a magnetic dipole corresponding to an orientation $w$. The second left column illustrate a magnetic quadrupole with principle axis aligned to the $z$-axis corresponding to a longitudinal alignment $X$. It is driven by the total intensity $S_0$ independent of the polarization state. The remaining three columns illustrate quadrupoles in the $x$-$y$-plane driven by linearly polarized light in the equatorial plane of the  Poincar{\'e} sphere. It corresponds to the $\Delta m=2$-coherence $\Phi$ in the atom representing a transverse alignment. Light polarized along the $x$-direction (centre column, pump polarization used in experiments) just drives the real part $u$ of the coherence represented by a quadrupole with one (north) of the principal axes along the $x$-direction and the other along $y$. Light polarized at 45$^\circ$ drives the imaginary part $v$ with the north principal axis oriented at 45$^\circ$. In general, light with a phase $\phi_L$ between $\sigma$ components (second to right column) drives a quadrupole with north principle axis at the polarization angle $\phi_Q =\phi_p=\phi_L/2$ (conventional optics notation, $(\phi_L -\pi)/2$ in convention used in our theoretical description).}
\end{figure}

Linearly polarised light, being a superposition of $\sigma_+$ and $\sigma_-$ light will drive various $\Lambda$- and $V$-schemes creating $\Delta m =2$-coherences. It is convenient to take the wavevector of the forward beam as the quantization axis (positive $z$-axis). In this representation a longitudinal magnetic field $\Omega_z$ will split the Zeeman substates energetically, whereas transverse fields $\Omega_x$, $\Omega_y$ will induce coherences (only one example is shown in Fig.~\ref{fig:levelscheme}). Obviously, the level scheme is quite complex. Fortunately, there is another representation in terms of irreducible tensor components (e.g.\ \cite{omont77}). The tensor $\rho^0_0$ is simply the total population of the ground state, assumed to be one in the following. The tensor of rank 1 denote dipole states. $\rho_0^1$ denotes the orientation or longitudinal magnetization of the ensemble (Fig.~\ref{fig:moments}, lower row, leftmost structure). $\rho_{\pm1}^1$ is related to the transverse magnetization. Tensors of rank two denote quadrupole states. $\rho_{0}^2$ is the longitudinal alignment displayed in the second left panel of the lower row of Fig.~\ref{fig:moments}, $\rho_{\pm 2}^2$ are transverse alignments. Superpositions of these are displayed in the three right panels of the lower row of Fig.~\ref{fig:moments}. In addition there are tensor moments of rank 3 but they do not provide back action onto the field and we are going to neglect them in the analysis.

\nointerlineskip

All the tensor components up to rank 2 are also present in the simpler $F=1\to F'=2$-transition depicted in Fig.~\ref{fig:levelscheme}\textbf{(b)}. Expressed in density matrix components, they are given by
\begin{equation} \label{eq:w}
w:= \rho_{11}-\rho_{-1 -1}\sim \rho_0^1,
\end{equation}
denoting the orientation (longitudinal magnetization),
\begin{equation} \label{eq:x}
X:= \rho_{11}+\rho_{-1 -1}-2\rho_{00}\sim \rho_0^2,
\end{equation}
denoting the longitudinal alignment,
\begin{eqnarray}
\Phi & := & 2\rho_{1 -1} = u+iv \sim \rho_{-2}^2,\label{eq:Phi}\\
u & := & \rho_{-1 1}+\rho_{1 -1} \label{eq:u},\\
v &:= & i(\rho_{-1 1}-\rho_{1 -1}) \label{eq:v},
\end{eqnarray}
denoting the $\Delta m=2$ ground state coherence and its real and imaginary parts (transverse alignment) respectively. Representations of these are displayed in the three right panels of the lower row of Fig.~\ref{fig:moments}.

The $F=1/2\to F'=3/2$-transition depicted in Fig.~\ref{fig:levelscheme}\textbf{(c)} allows only for an orientation (rank 1).  It is however instructive to illustrate the back action of the atoms on the $\sigma$ components in the light field. A negative orientation will enhance the optical density and phase shift for $\sigma_-$ light and decrease it for $\sigma_+$ light and vice versa. It is important that in contrast to the $J=1/2\to J'=1/2$-transition on the $D_1$-line used in \cite{grynberg94a,ackemann95b, ackemann01b} the trapping state is not `dark' but `bright' and as optical density is increased and not reduced with increasing intensity, the nonlinearity is self-focusing for red detuning and self-defocusing for blue detuning, opposite to the electronic transition and the optical pumping nonlinearity on $D_1$-lines. The $F=1/2\to F'=3/2$-transition was used in \cite{schmittberger16} to describe some vectorial aspects of self-organization in counterpropagating beams.

For a theoretical treatment, we follow the recipe in \cite{mitschke86} and write down the Liouville equation for the density matrix describing the semiclassical interaction of the light field with the atoms and add relaxation terms. Then the degrees of freedom of the excited state are adiabatically eliminated. The results are inserted into the equation of motion of the ground state variables and from then on it is assumed that all population resides in the ground state. Explanations and the equations can be found in \cite{kresic18,labeyrie18} and the related supplementary material. Further details are in \cite{kresic17t}. The back action is given by
\begin{equation}\label{eq:chi_mag}
\frac{\partial} {\partial z} E_\pm=i \chi_\pm \frac {k}{2} \left[ \left(1\pm\frac{3}{4}w+\frac{1}{20}X\right)E_\pm+
\frac{3}{20}(u\mp iv)E_\mp\right],
\end{equation}
where the linear susceptibility $\chi_\pm$ is
\begin{equation} \label{eq:chi_maglin}
\chi_\pm=\frac{b_0}{k L}\frac{i+\bar{\Delta} \mp \Omega'_z}{1+(\bar{\Delta} \mp \Omega'_z)^2}.
\end{equation}
Here, $\bar{\Delta}$ and $\Omega'_z$ are normalized to the decay rate of the optical polarization $\Gamma_2=\Gamma/2$ to simplify notation. The $\pm$ in front of the $w$ term expresses the change of optical density from a ground state orientation discussed above and will be discussed in detail in Sec.~\ref{sec:dipole}. The longitudinal alignment will not induce a polarization instability but change optical density. There are indications that this can lead to self-organization under certain conditions \cite{labeyrie18} but we will not discuss it in detail here. The real part $u$ of the coherence also will not create a polarization instability, but the imaginary part $v$ of the coherence provides an additional novel possibility for a polarization instability not present in a rank 1 system ($F=1/2\to F'=3/2$ or $J=1/2\to J'=1/2$). Note that the coupling between $\sigma_+$ and $\sigma_-$ mediated by $v$ is quite different from the one mediated by $w$. We will discuss this in Sec.~\ref{sec:quadro}.

\subsection{Dipolar structures}\label{sec:dipole}
\begin{figure}[h]
\centering
\includegraphics[width=8cm]{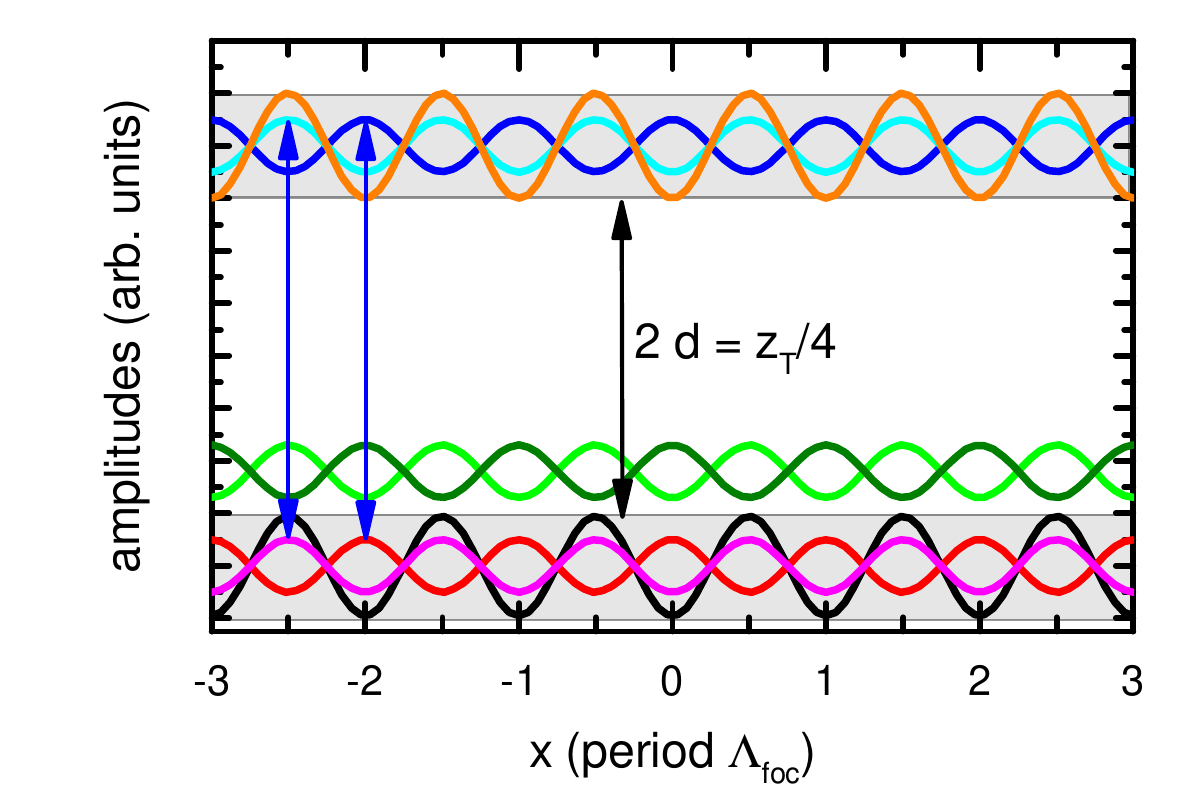}
\caption{\label{fig:feedback_mag} Illustration of feedback loop for single mirror feedback for magnetic dipole interactions (see caption of Fig.~\ref{fig:feedback_electronic} for general explanation).  Black line: orientation $w$. For red detuning: red (magenta) line: refractive index for $\sigma_-$ ($\sigma_+$) light, light (dark) green line: phase modulation of transmitted $\sigma_-$ ($\sigma_+$) beam, dark (light) blue: intensity of $\sigma_-$ ($\sigma_+$) beam after a quarter of the Talbot distance, orange: difference pump rate $D$.}
\end{figure}

The orientation in the atomic cloud is driven by the helicity of the driving light beam  linked to the Stokes parameter $S_3$ (see leftmost column of Fig.~\ref{fig:moments}). It is driven by the difference $D:=P_+ - P_-$ of the pump rates of $P_+$, $P_-$ of the $\sigma_{\pm}$ polarization components. The Stokes parameter is given by $S_3=D/S$, where $S=P_+ + P_-$ represents the sum pump rate or Stokes parameter $S_0$. A simplified equation of motion is given by \cite{kresic18}
\begin{equation} \label{eq:w_dyn}
\frac{d}{dt}w = - r + \frac{5}{18} D - \frac{1}{6}{S} \,+\,\mbox{coupling to other multipoles},
\end{equation}
where $r$ represents an effective relaxation constant ($r\approx 2.8\times 10^3$s$^{-1}$  for $\Lambda \approx 100$~$\mu$m and $T \approx 200$~$\mu$K for the experiment at INPHYNI; $r\approx 4.4\times 10^3$s$^{-1}$ for $\Lambda \approx 50$~$\mu$m and $T \approx 120$~$\mu$K for the experiment at Strathclyde). The second term represent the driving, the third term saturation. The feedback loop leading to a spontaneous magnetization is depicted in Fig.~\ref{fig:feedback_mag}. The black line denotes a modulation of the orientation $w$. The red and magenta lines denote the resulting refractive index distributions experienced by $\sigma_{\pm}$ components for red detuning. The are shifted by half a period due to the $\pm$ sign in Eq.~(\ref{eq:chi_mag}). After a quarter of the Talbot distance the intensities distributions for the $\sigma_\pm$ components are shifted by half a period and each of them is aligned with the original refractive index distribution as required for a self-focusing situation. More importantly, the difference pump rate is spatially aligned with the orientation distribution and can give positive feedback for spontaneous growth. An optical spin pattern is sustaining the atomic spin pattern and vice versa.
The experiment is performed at red detuning to the $F=2\to F'=3$ $D_2$ line of $^{87}$Rb. Care is taken to create a linear input polarization of good quality and to avoid depolarization at the cell windows as far as possible. A small deviation of the angle of incidence by a few degrees from the normal needs to be accepted to avoid reflections from the cell windows seeding structures. However, this has no apparent detrimental effects on the structures.

\begin{figure}[h]
\centering
\includegraphics[width=13cm]{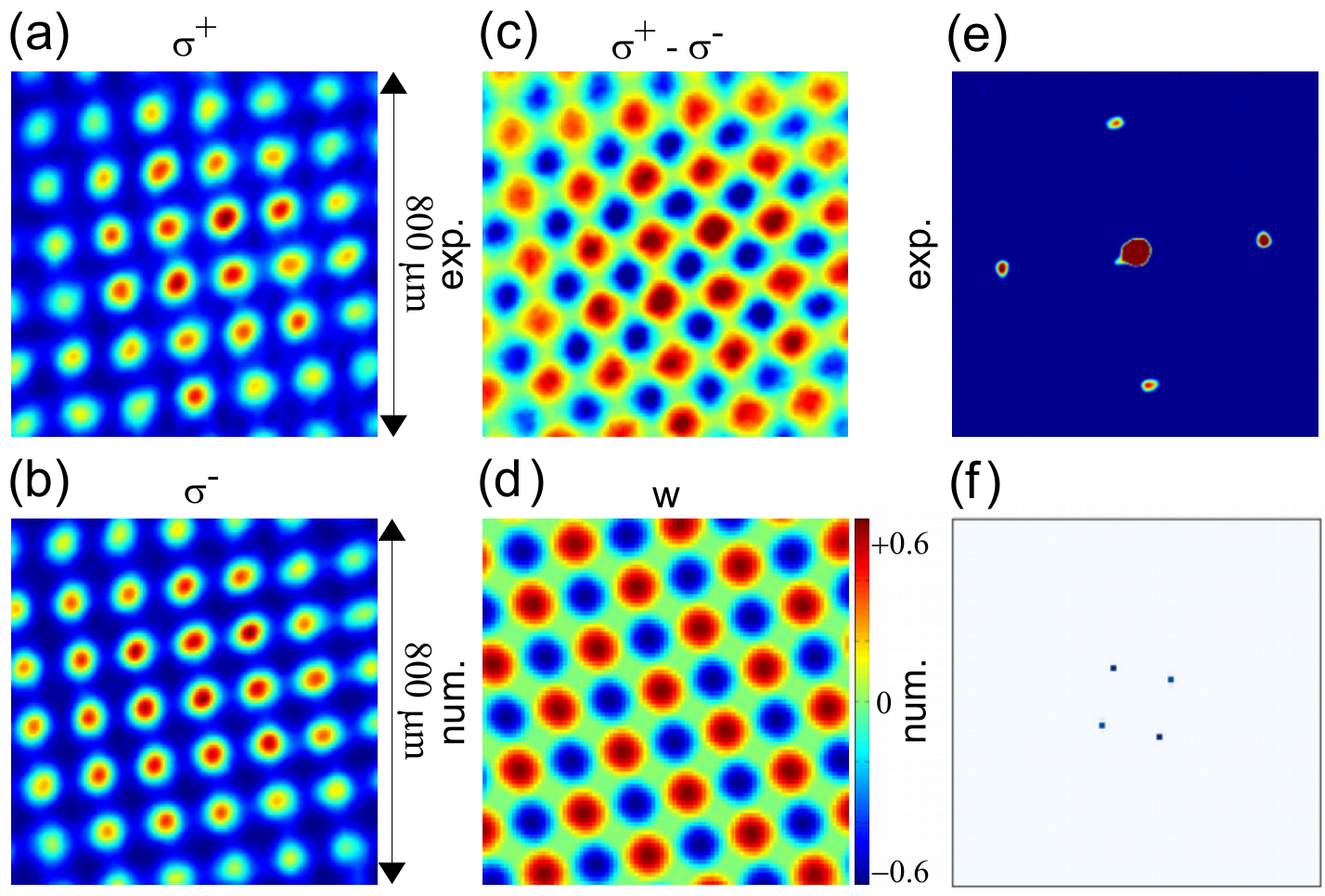}
\caption{\label{fig:squares}Spontaneous antiferromagnetic ordering for $B_x = B_y = B_z =0$.
\textbf{(a)}, \textbf{(b)}: NF intensity pattern of the $\sigma^+$ and $\sigma^-$ components. \textbf{(c)}, NF intensity difference of the $\sigma^+$ and $\sigma^-$ components. \textbf{(d)} Spatial structure of the orientation $w$ obtained from numerical simulation. \textbf{(e)} example for FF intensity distribution experimentally observed (not the same realization as in \textbf{(a)} -\textbf{(c)}).  \textbf{(e)} Numerically obtained Fourier spectrum of $w$. Parameter of experiment: $b_0=80$, $\delta = -8 \Gamma$,   $I = 10$~mW/cm$^2$, $d = -20$~mm. (subpanels \textbf{(a)}-\textbf{(d)} adapted from \cite{kresic18})}
\end{figure}

The typical observation in zero magnetic field is illustrated in Fig.~\ref{fig:squares}. The input field being a superposition of equal amounts of $\sigma_\pm$ light breaks up into patches of square symmetry in which locally one of the $\sigma$ components is dominating (Fig.~\ref{fig:squares}\textbf{(a)} and \textbf{(b)}). The two lattices are interlaced, i.e.\ shifted by half a period. A visualization of the orientation in the cloud is possible by subtracting the images of the $\sigma$ components from each other (this imaging becomes exact in the first order approximation in the susceptibility to obtain the transmitted field, see (\ref{eq:chi_mag})). Fig.~\ref{fig:squares}\textbf{(c)} shows the resulting square pattern of atomic orientation. The magnetization is modulated from positive to negative on zero offset. The amplitude of modulation is (within experimental imperfections) equal in the positive and negative direction, i.e.\ this is an anti-ferromagnetic state. This result is in excellent agreement with a numerical calculation of the orientation $w$ shown in Fig.~\ref{fig:squares}\textbf{(d)}. As expected from the discussion of Fig.~\ref{fig:feedback_mag}, optical and atomic spin patterns sustain each other.

The threshold for this spontaneous magnetic ordering is very low.  For the INPHYNI experiment, $b_0=80$, $\delta=-8.6\Gamma$, the linear phase shift is $|\phi_{lin}|\approx 2.3$ and the threshold is about 2~mW/cm$^2$. The linear stability analysis is described in \cite{kresic18} and predicts a threshold of $I_{\rm th} \approx 0.1$~mW/cm$^2$. The difference can be easily explained by stray magnetic fields. The optical density in the Strathclyde experiment is measured to be $b_0=27$, giving nominally $|\phi_{lin}|=0.96$ at $\delta=-7\Gamma$, but we find a more robust agreement between theory and experiment assuming $b_0=30$ \cite{kresic19}, giving  $|\phi_{lin}|\approx 1.1$. Thresholds predicted are on the order of a few mW/cm$^2$ depending on exact parameters. Measured thresholds are between just smaller than 1~mW/cm$^2$ and 5~mW/cm$^2$. The variations can be explained by stray magnetic fields and fluctuations of atom number. An input intensity of 1~mW/cm$^2$ corresponds to an electronic saturation level of only about $s\approx 10^{-3}$, so thresholds are indeed very low and the description in terms of ground state dynamics well satisfied.

It should be noted that the dynamics can be qualitatively understood by looking at the orientation $w$ only but that other multipoles are important on a quantitative level, especially for low optical densities \cite{kresic18}. Without relaxation, $r=0$, the pump threshold is zero but still a finite linear phase shift is needed for the instability. This can be estimated to be about $|\phi_{lin}|\approx 0.8$ based on $w$ alone and  $|\phi_{lin}|\approx 0.9$ including some effects of higher multipoles. Hence the Strathclyde experiment with $b_0\approx 27-30$ is operating quite at the limit. Experiments varying the optical density at INPHYNI yielded a threshold in optical density of 25 for a detuning of 12$\Gamma$ (see Fig.~\ref{fig:OD_threshold}). This corresponds only to  $|\phi_{lin}|\approx 0.5$, of the same magnitude, but smaller. Further more systematic experimental work including a careful analysis of the uncertainty of the determination of the optical density and theoretical work including all atomic degress of freedom is necessary to establish an accurate limit. However, it is clear that the magnetic ordering poses much less stringent requirements on optical density than the electronic one. At Strathclyde,  the apparatus was recently upgraded and it is planned to increase atom number by about a factor of 10 and the optical density by a factor of 2 \cite{costa-boquete20}.

As a final remark, we mention that as the homogeneous solution is $w_h=0$, there is no upper threshold and the instability persists for arbitrarily high saturation in the model as the saturating term related to the homogeneous solution discussed for Eq.~(\ref{eq:eta}) is not present. For the experimental realization, one is limited to the saturation range in which the dynamics is accurately described by the confinement to optical pumping in the ground state, of course.

If a transverse magnetic field of sufficient strength is applied, the magnetic structuring is suppressed \cite{kresic18}. This is easily understood in classical terms as a magnetic field in $x$ ($y$) direction will induce a precession in the $y-z$ ($x-z$) plane and thus the longitudinal component will average out. In the quantum mechanical description, the transverse field stimulates coherent coupling (tunneling) between spin states thus suppressing ordering as described by the transverse (also referred to as the quantum) Ising model \cite{sachdev11,dutta15,strecka15}. The transverse Ising model displays, even at zero temperature, a quantum phase transition between the ordered antiferromagnetic state and a paramagnetic state in which all spins are aligned to the transverse field.  In the feedback-induced ordered state the transverse magnetic coupling induces a phenomenologically similar transition from the antiferromagnet to a state without structure. Furthermore the scaling properties agree. The magnetization decays as $\sqrt{1-(B/B_{\rm crit})^2}$ in agreement with the transverse Ising model \cite{strecka15} and the magnetic order parameter (staggered magnetization for the anti-ferromagnetic case) scales like a square-root with the distance of the pump parameter to the critical point, as expected for the Ising model in the mean field limit \cite{strecka15}. The main source of disorder to be overcome is the residual thermal motion of the atoms leading to a wash-out of the spin pattern and thus serving as a kind of effective spin temperature counteracting the magnetic ordering. At constant pump power once can think of sweeping the relaxation parameter $r$ as a measure of effective temperature to induce the phase transition. We will discuss further connections to the Ising model in the two next sections.

\subsection{Hexagon formation and inversion symmetry}\label{sec:sym}
If a magnetic field of the order of a few tens of mG (the exact value depends on other parameters as intensity), the squares give way to hexagonal patterns \cite{kresic18,labeyrie18,kresic19}. For $B_z >0$, these are hexagonally coordinated peaks (like the ones displayed in Fig.~\ref{fig:symbreaking}) for the $\sigma_+$ component and the orientation $w$ and honeycombs in the $\sigma_-$ component. These are ferrimagnetic states, i.e.\ the structures have a dominantly positive orientation (Fig.~3 of \cite{kresic18})  but are not completely ferromagnetic as the magnetization dips below zero between the dominant peaks. For $B_z <0$, all roles are reversed. This also agrees with the expectation from the Ising model, a longitudinal bias field will induce a preferred magnetization. In a normal magnet, this is due to the dipole energy, here due to coupling to the coherent nonlinear Faraday effect and incoherent linear  and nonlinear Faraday effects \cite{labeyrie18,kresic19}. The former is related to a coupling of the orientation to the $\Delta m=2$-coherence $\Phi$ \cite{kresic19,wojciechowski10}, the latter two due to the presence of the longitudinal magnetic fields in the detuning terms of the susceptibility (Eq.~(\ref{eq:chi_mag}), see also \cite{labeyrie01}) and pump rates \cite{labeyrie18}.

Inspecting the equation of motion (\ref{eq:w_dyn}), one realizes that it is inversion symmetric. A change of sign of the orientation $w\to -w$ can be easily shown to lead to an exchange between $P_+ \leftrightarrow P_-$ or $D\to -D$. This symmetry is broken by the presence of the longitudinal magnetic field as it breaks the symmetry between the pump rates: The same intensity will lead to a different pump rate as the detuning value changes differently for the $\sigma_\pm$ components (incoherent nonlinear Faraday effect). It turns out that the symmetry breaking is enhanced in a $J=1$ ground state compared to a $J=1/2$  due to the coherent Faraday effect based on $\Delta m=2$-coherences \cite{kresic19}. The transition to hexagonal structures when the inversion symmetry is broken can be understood using a very general symmetry argument \cite{busse78,cross93,gollwitzer10}. The competition between different critical modes with complex amplitude $A_i$ with the critical wavenumber can be phrased in a Ginzburg-Landau type equation as
\begin{equation}\label{eq:GLE}
 \frac{d A_i}{dt} = \eta A_i + \eta_2 A_j^\star  A_k^\star \, - \,\eta_{3,c}\sum_{j\neq k} |A_j|^2 A_i \, -\, \eta_{3,s} |A_i|^2 A_i ,
\end{equation}
where $\eta$ is the linear growth rate obtained by a linear stability analysis (e.g.\ Eq.~(\ref{eq:eta}), $\eta_2$ a quadratic coupling coefficient ($i\neq j \neq k$ in the quadratic term) and $\eta_{3,s}$, $\eta_{3,c}$ cubic self- and cross-coupling coefficients saturating the growth of the instability when positive. These latter terms are always phase-matched, i.e.\ the sum of the involved transverse lattice wave vectors yields a vector with the wavenumber of the lattice. The second order term is only different from zero if the system has no inversion symmetry, i.e.\ $B_z \neq 0$, but will dominate over the third order term as long as the amplitudes of the unstable modes are small enough close to threshold. This favours hexagonal patterns as the sum of two lattice wavevectors will yield a vector with the lattice wavenumber only for an angle of 120$^\circ$. The quadratic coupling coefficient for the system under study was explicitly calculated in \cite{kresic19} and confirms the symmetry breaking properties via the coherent (for small $B_z$) and incoherent (for large $B_z$) Faraday effect.

The scenario described here has not been observed in previous experiments in spin-1/2 sodium or Rb vapours. The coherent Faraday effect (for small $B_z$) does not exist as there is no tensor of rank 2 in the $J=1/2$ ground state. The inversion symmetry breaking due to the incoherent Faraday effects (expected for large enough $B_z$) had not been sufficient for a sizable effect in hot vapours as for moderate magnetic fields in the Gauss range the change of detuning is negligible for the conditions of high pressure broadening used in  \cite{aumann97,aumann99t,aumann00}. In the Rb experiments without buffer gas the aspect ratio was not sufficient to observe hexagons or squares and the flower-like patterns observed in \cite{grynberg94} stem from the circular symmetry imposed by the Gaussian input beam \cite{leberre95b}. Symmetry breaking can be induced by slightly perturbing the linear input state and injecting light with a small polarization ellipticity, i.e.\ difference pump rate. This was theoretically predicted in \cite{scroggie95t,leberre95b,scroggie96} numerically, in \cite{scroggie95t,scroggie96,aumann99t} analytically (in models of different complexity) and experimentally verified in \cite{aumann97}. Reviews are in \cite{lange98a,aumann04}. A symmetry breaking transition vs.\ the longitudinal field has been obtained in hot $J=1/2$ atoms only if an additional transverse magnetic field is present \cite{aumann00,logvin00}, but relies on light-shifts changing the ground state degeneracy.

For an inversion symmetric system with $\eta_2=0$, pattern selection takes place in third order and is non-generic but depends on details of the relative strength of the self- and cross-coupling coefficients $\eta_{3,s}$, $\eta_{3,c}$ and the angle-dependence of the cross-coupling coefficients $\eta_{3,c}$ \cite{busse78,cross93,gollwitzer10}. Calculations for $J=1/2$-atoms confirmed that in wide parameter regimes squares should be the preferred state selected for a self-focusing situation \cite{scroggie95t,leberre95b,scroggie96,leduc96,aumann99t}. As there is emerging interest in 2D quasiperiodic optical lattices in the field of quantum simulation \cite{viebahn19,mivehvar19}, we mention that the selected state for a self-defocusing situation are quasiperiodic structures with an eight-fold rotational symmetry \cite{leduc96,leduc96a}. This was experimentally and analytically (using a more complex model) verified in \cite{aumann99t,aumann02}. Quasiperiodic states with twelve-fold rotational symmetry have been predicted for a quasi-scalar situation using optical pumping in a $J=1/2$ ground state in \cite{leberre98} and experimentally observed in \cite{herrero99}. It should be noted that emergence and rotational symmetry of the structures are self-organized in these studies. The rotational symmetry of quasiperiodic structures discussed and observed in \cite{vorontsov93,pampaloni95} is imposed by a field rotation in the feedback loop, restricting the continuous rotational symmetry to a discrete one. Superlattice states were observed and analyzed in \cite{grossewesthoff03}.

For the experimental situations with dominating optomechanical and electronic nonlinearity, the inversion symmetry is broken from the start and hence hexagons are the structures obtained at threshold as indicated in Figs.~\ref{fig:Talbot}, \ref{fig:symbreaking}. For a Kerr nonlinearity this was explicitly calculated in \cite{dalessandro92}, for an optomechanical model including velocity damping via an external molasses in \cite{baio21}. Interestingly, in the latter model, inversion symmetry is reestablished at a critical coupling strength and a transition from honeycomb to hexagonal via stripe structures is obtained in the atomic density changing the linear phase shift \cite{baio21}. This confirms that the optomechanical system is richer than just being an `artificial Kerr'-medium \cite{ashkin82} due to the transport character of the nonlinearity. A transition between hexagonal and honeycomb structures has been predicted also for supersolids in dipolar quantum gases \cite{zhang19,zhang21}. For Hamiltonian systems as in \cite{zhang19} selection principles like in Eq.~\ref{eq:GLE} arise from energy functionals. In the case of the open, driven systems discussed here it arises from a generalized free energy functional \cite{cross93}. Due to this analogy, these considerations are also of high interest to understand periodic as well as quasiperiodic symmetries of new quantum phases in the field of quantum simulation of many-body states.


\subsection{Quadrupole structures} \label{sec:quadro}
Apart from helicity described by the $S_3$ Stokes parameter, light has the polarization direction as degree of freedom on the Poincar{\'e} sphere. It is described by the Stokes parameters $S_1$ and $S_2$ determining the azimuthal angle in the equatorial plane (three rightmost columns in Fig.~\ref{fig:moments}). Taking the field as a superposition of $\Omega_+$ and $\Omega_- \exp{(-\phi_L)}$ with $\Omega_+$, $\Omega_-$ real and $\phi_L$ denoting the phase difference, the polarization angle $\phi_p$ for linearly polarized light (or the major principal axis for elliptically polarized light) is given by
\begin{equation}\label{eq:polangle}
\phi_p= \frac{\phi_L-\pi}{2}.
\end{equation}
This slightly cumbersome relation is due to the peculiarities of the notation used in \cite{kresic18} and useful to describe the tensorial light-matter interaction with the quantization axis in positive $z$-direction, but in conventional optical notation (looking towards the beam, opposite to the wavevector) $\phi_L$ is directly the angle on the Poincar{\'e} sphere and $\phi_p = \phi_L/2$ the polarization direction (as indicated for simplicity in Fig.~\ref{fig:moments}). The $\Delta m=2$-coherences $u$, $v$ are then driven by the Raman pump rates
\begin{eqnarray}
P_{\Lambda+} & = & \frac{2\Omega'_+ \Omega'_-}{1+\bar{\Delta}^2} \,  \cos{\phi_L} \sim - S_1,\\
P_{\Lambda-} & = & -\frac{2 \Omega'_+ \Omega'_-}{1+\bar{\Delta}^2} \, \sin{\phi_L} \sim - S_2,
\end{eqnarray}
like
\begin{eqnarray}
\dot{u} & = & -\Gamma_c u \, +\,  \frac{5}{18}P_{\Lambda+} \, + \, \mbox{coupling to other multipoles} ,\\
\dot{v} & = & -\Gamma_c v \, -\, \frac{5}{18}P_{\Lambda-} \, + \, \mbox{coupling to other multipoles}.\\
\end{eqnarray}
Here $\Gamma_c$ is a relaxation rate depending on $r$ and sum pump rates and $\Omega'_+$, $\Omega'_-$ are the Rabi frequencies of the $\sigma_\pm$ components normalized to $\Gamma /2$. $P_{\Lambda+}$ is the negative of the Stokes parameter $S_1$, $P_{\Lambda-}$ the negative of $S_2$ (again, the minus signs are a result of the convention chosen). The stationary solutions are
\begin{eqnarray}
  u &\sim & \frac{2\Omega'_+ \Omega'_-}{1+\Delta^2} \,  \cos{\phi_L} ,
  \\
  v &\sim & - \frac{2\Omega'_+ \Omega'_-}{1+\Delta^2}\,  \sin{\phi_L}\\
  \Phi & \sim & \frac{2\Omega'_+ \Omega'_-}{1+\Delta^2} \, e^{-i\phi_L} .
\end{eqnarray}
This shows that the phase $\phi_L$ between the circular polarization components determines the phase of the coherence $\Phi$. The direction of the principal axis of the quadrupole $\phi_Q$ is linked to the phase of the coherence via an equation like (\ref{eq:polangle}). Hence  the polarization direction of the light is directly controlling the direction of the quadrupole, $\phi_L=\phi_Q$, which is of course expected from symmetry arguments. For example, in the notation used the $x$-polarized input beam corresponds to $\phi_L=\pi$ and $\phi_p=0$.
It pumps, $u\neq0$, $v=0$, i.e.\ the cones of the resulting quadrupole (centre representation of $u$ in Fig.\ref{fig:moments}) are directed along the $x$-axis, the orthogonal cones along $y$.  $\phi_L=\pi/2$ corresponds to light polarized at 45$^\circ$ to the $x$-axis. It pumps  $u=0$, $v\neq 0$, i.e.\ the cones of the resulting quadrupole (rightmost representation of $v$ in  Fig.\ref{fig:moments}) are directed at $45^\circ$  along the bisections of the  $x$-axis and $y$-axis. In between, there is a smooth transition stemming from the form of the spherical harmonic function (second to right representation of $|\Phi|$ in Fig.\ref{fig:moments}).

\begin{figure}[h]
\centering
\includegraphics[width=10cm]{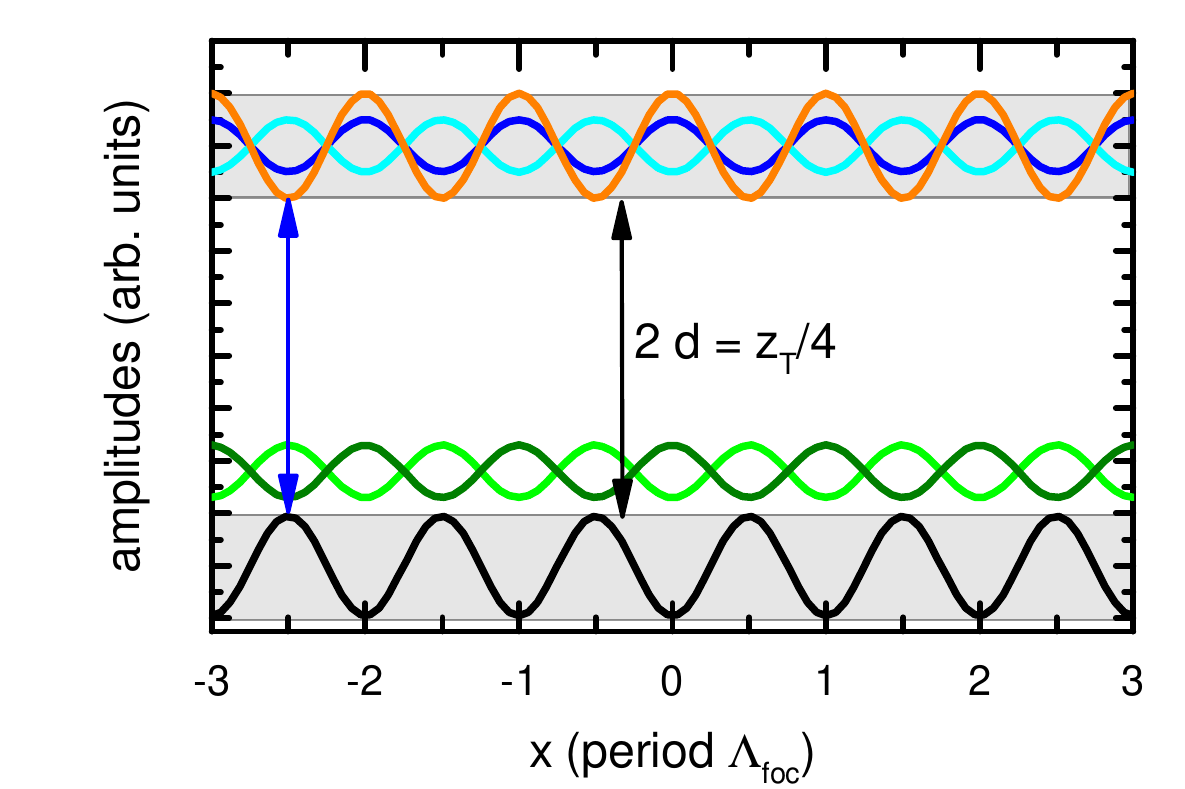}
\caption{\label{fig:feedback_coherence} Illustration of feedback loop for single mirror feedback for quadrupole interactions (see caption of Fig.~\ref{fig:feedback_electronic} for general explanation).  Black line: modulation of imaginary part of coherences $v$. For red detuning: light (dark) green line: amplitude modulation of transmitted beam for $\sigma_-$ ($\sigma_+$), light (dark) blue: phase of $\sigma_-$ ($\sigma_+$) after a quarter of the Talbot distance, orange line: relative phase between circular polarization components $\phi_L$ and modulated polarization direction.}
\end{figure}

As indicated, an input beam linearly polarized in the $x$-direction will pump only the real part $u$ of the coherence. However, an instability is possible from a fluctuation in the imaginary part $v$ (black line in Fig.~\ref{fig:feedback_coherence}). In the large detuning limit of a real linear susceptibility, the imaginary prefactor of $v$ will provide an imbalance in transmission for the $\sigma$-components. For positive $v$ and red detuning, the $\sigma_-$ component is amplified, the $\sigma_+$ component is attenuated. This phenomenon is of course well known for coherently driven $\Lambda$ transitions and related to electromagnetically induced transparency \cite{fleischhauer05}. Differently from the situations discussed before, the transmitted fields after the medium are now amplitude modulated.  This amplitude modulation is converted to a phase modulation by the diffractive dephasing in the feedback loop. This implies that the relative phase $\phi_L$ and thus the polarization direction $\phi_p$ is modulated. At a quarter of the Talbot distance $\phi_L$ is anti-phased to $v$ and as $v \sim -\sin{\phi_L}$, there is positive feedback. For constant $u$, the modulation of $v$ simply transfers into a modulation of the quadrupole direction $\phi_Q$. Hence a modulation of polarization angle $\phi_p$ or phase modulation $\phi_L$ sustains a modulation of the phase of the coherence and quadrupole direction in the medium and vice versa.

\begin{figure}[h]
\includegraphics[width=15cm]{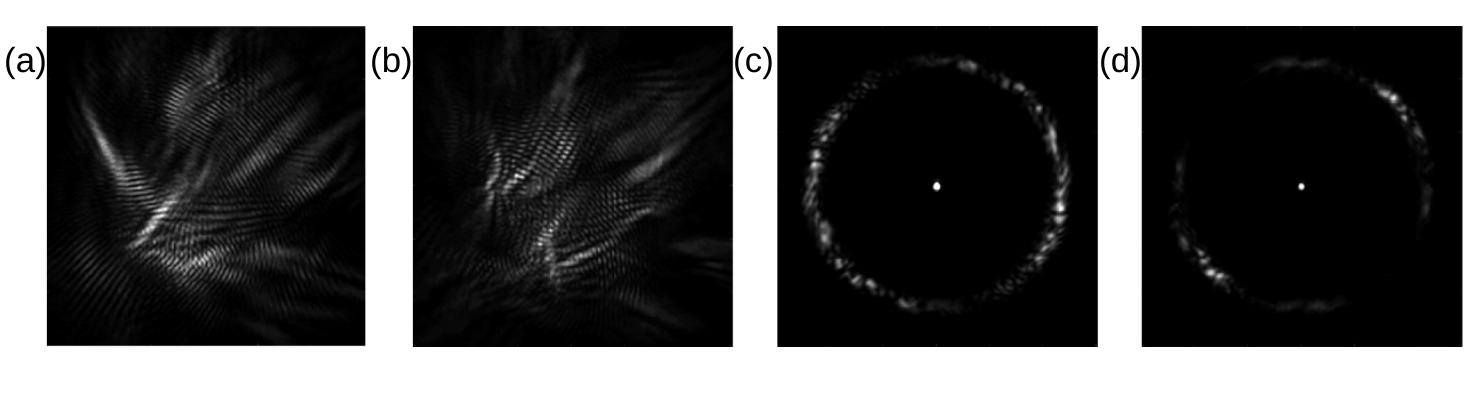}
\caption{\label{fig:flash_exp}
Example for coherence-based structure detected in channel perpendicular to input polarization. \textbf{(a)}, \textbf{(b)}: Near field intensity structures. \textbf{(c)}, \textbf{(d)}: Far field intensity displaying Fourier spectrum (not obtained in same realizations as in \textbf{(a)}, \textbf{(b)}).  Parameters: pump intensity 7 mW/cm$^2$, detuning $\Delta = -12 \Gamma$, $b_0=110$. The field of view is 4.4 mm for the NF  and 16 mrad for the FF images.}
\end{figure}

We relate this feedback mechanism to a peculiar phase appearing at sufficiently high optical density, if the transverse field component parallel to the input polarization is increased beyond the critical field strength at which the dipolar square patterns disappear \cite{labeyrie18}. These structures have a locally stripe-like or rhombic appearance (Fig.~\ref{fig:flash_exp}\textbf{(a)}, \textbf{(b)}) but are usually highly disordered and contain many defects. Typically, nearly the whole critical ring in Fourier space is excited (Fig.~\ref{fig:flash_exp}\textbf{(c)}), although sometimes one direction of wavevectors is dominating (Fig.~\ref{fig:flash_exp}\textbf{(d)}). Injecting a circularly polarized probe beam after the pump beam is switched off,  a ring in Fourier space is detected in the opposite circular polarization for these patterns, but not for square patterns \cite{labeyrie18}. This provides evidence that the coherences are important in forming and sustaining these structures.

Coherence-based structures are not observed in the experiment at Strathclyde as the optical density achieved there is too low. Depending on parameters, in particular optical density and intensity, there is a gap in the existence range of squares and the irregular, coherence-based structures vs.\ $B_x$, or a direct transition from squares to irregular, coherence-based structures. We stress that the threshold for the formation of these structures is very low. Values smaller than 1 mW/cm$^2$ are obtained without a systematic investigation on the possibility of optimization. This corresponds to electronic saturation parameters of the $\sigma$ transitions of less than $10^{-3}$.

\begin{figure}[h]
\includegraphics[width=15cm]{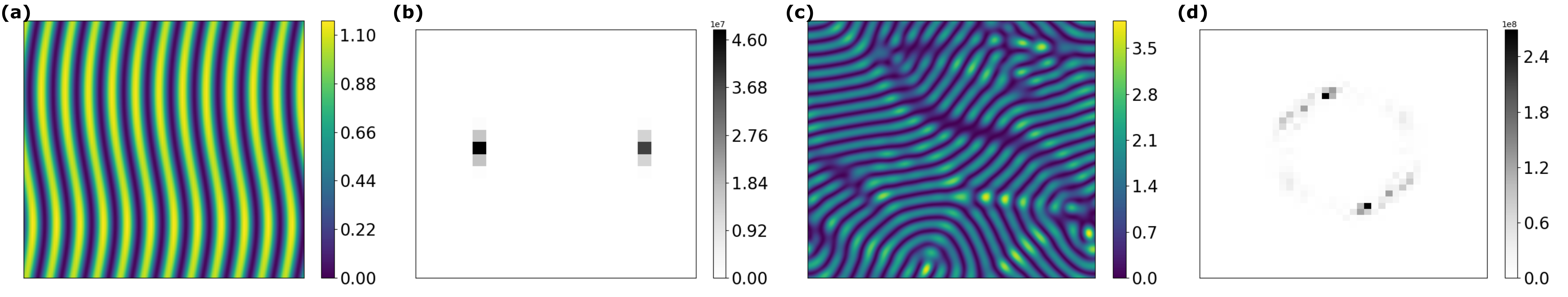}
\caption{\label{fig:flash_sim} Examples for numerically obtained structures. \textbf{(a)}, \textbf{(c)}: Near field intensity structures. \textbf{(b)}, \textbf{(d)}: Corresponding far field intensity displaying Fourier spectra. Parameters $B_x = 0.76$~G, $B_y=B_z = 0$, $b_0= 130$, $\Delta = -12 \Gamma$, pump intensity 6.8 mW/cm$^2$. The difference between the two realization is the size of the integration domain relative to the lattice period.}
\end{figure}

\begin{figure}[h]
\centering
\includegraphics[width=12cm]{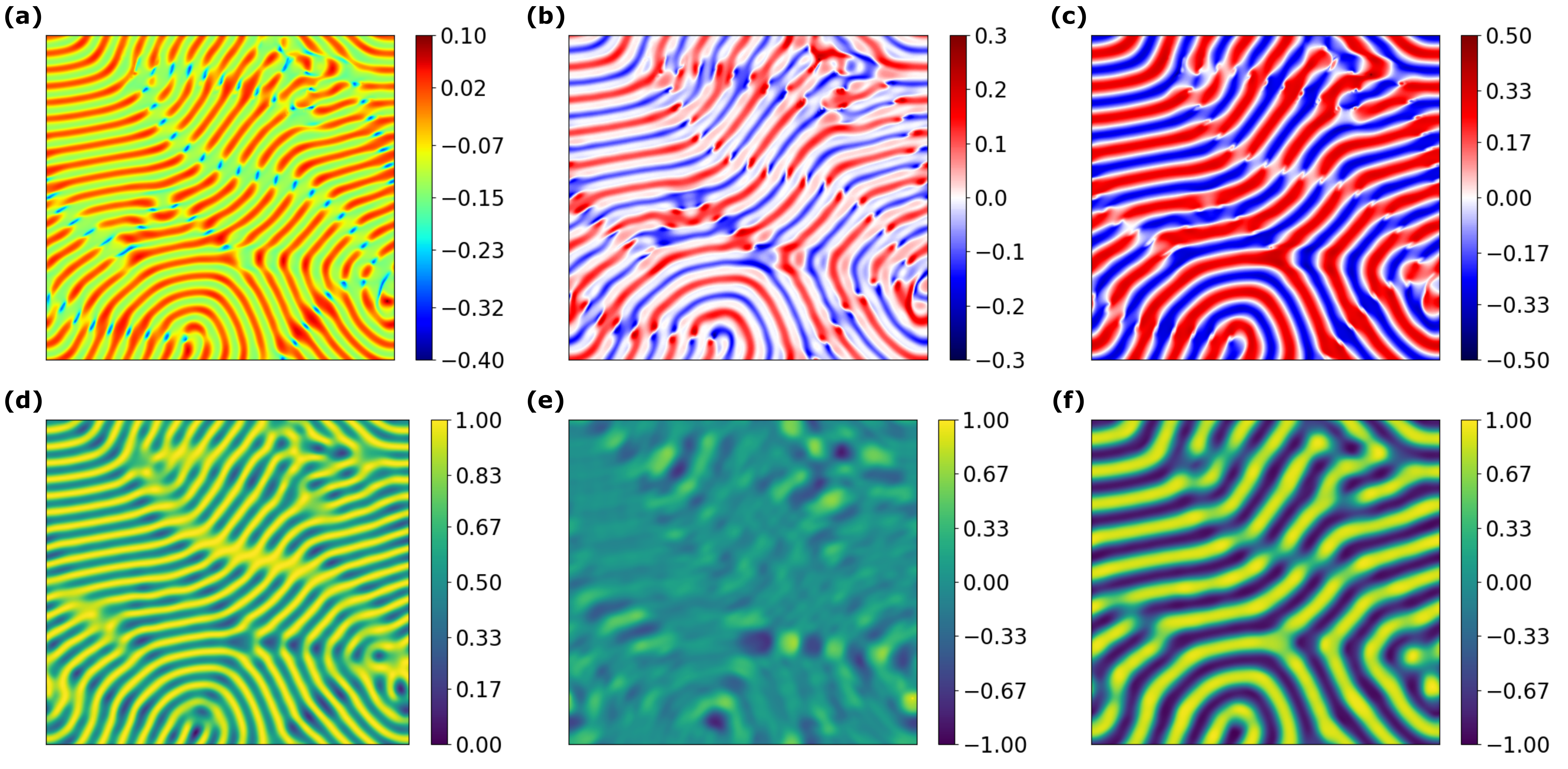}
\caption{\label{fig:flash_atomicStokes} Upper row: Spatial distribution of atomic variables corresponding to the structures in Fig.~\ref{fig:flash_sim} \textbf{(c)}:  \textbf{(a)} real part of coherence $u$,  \textbf{(b)} imaginary part of coherence $v$, \textbf{(c)} orientation $w$, Lower row: Stokes parameters of backward beam, \textbf{(d)} $S_1$, \textbf{(e)} $S_2$, \textbf{(f)} $S_3$.  }
\end{figure}

These structures are reproduced in numerical simulations. Examples are shown in Fig.~\ref{fig:flash_sim}. The principle structure are wavy stripes (Fig.~\ref{fig:flash_sim}\textbf{(a)}, \textbf{(b)}). In large enough domains these break up into patches of stripes with a lot of defects and domain boundaries (Fig.~\ref{fig:flash_sim}\textbf{(c)}, \textbf{(d)}). This matches the experimental observation. In the simulations also the atomic distributions are directly accessible. The imaginary part of the coherence $v$ (Fig.~\ref{fig:flash_atomicStokes}\textbf{(b)}) and the orientation $w$ (Fig.~\ref{fig:flash_atomicStokes}\textbf{(c)}) are strongly modulated at the critical lattice period. The modulation is around zero background and preserves inversion symmetry as expected without a longitudinal magnetic field. The real part of the coherence $u$ is modulated on a negative background. This modulation is at two times the critical lattice period and only secondary, driven by the modulations in $v$ and $w$. The lower row of Fig.~\ref{fig:flash_atomicStokes} illustrates that the Stokes parameters of the transmitted light can give some information on the atomic variables. $S_1$ provides a visualization of $u$ (Fig.~\ref{fig:flash_atomicStokes}\textbf{(d)}) and  $S_2$ of $u$ (Fig.~\ref{fig:flash_atomicStokes}\textbf{(e)}). The latter is however mainly sensitive to the defects although also a faint modulation is present in the domains of nearly regular stripe direction. We used the visualization of $w$ by $S_3$ (Fig.~\ref{fig:flash_atomicStokes}\textbf{(f)}) already in the previous section. We are currently investigating how to implement this experimentally as the disordered nature of the structure demands a careful overlap of the intensity measurements to obtain reliable Stokes parameters. It should be noted that there are coupling terms between the orientation $w$ and the coherence $v$. Hence the structures obtained are not purely of dipolar or quadrupolar in character but mixed. However, for the squares the dipolar component is much stronger than in the irregular phase. Details of these structures and the transition to squares via magnetic field and intensity are currently under investigation.

Some theoretical papers considered the possibility of coherence based self-organized states in cavities~\cite{oppo10,eslami14} but we are not aware of any previous experiment before Ref.~\cite{labeyrie18} demonstrating self-organized coherences. There has been a significant interest in exploring spatially structured coherences as image storage in quantum information technology \cite{vudyasetu08,shuker08,heinze13,ding13,radwell15}.
We also mention that advanced optical magnetometry using atomic vapours can be based on the dynamics of the alignment and not only on the orientation \cite{budker02,weis06,ingleby17,ingleby18}. As a final connection we point to spinor dynamics in Bose-Einstein condensates \cite{stenger98,sadler06,jacob12,pechkis13,anquez16}. Here states related to quadrupoles are often referred to as `nematics' and states  related to dipoles as `polar', as the former have a preferred orientation in space but not direction.

\section{Light-mediated atomic interaction}\label{sec:coupling}
We have studied in the previous sections how coupled instabilities of light and matter degrees of freedom lead to self-organization. In the optomechanical case, the atomic lattice creates an optical lattice via the scattering of the pump light into sidebands. The dipole potential of the optical lattice causes then the bunching of the atoms in a density structure. In the case of magnetic dipole ordering, an atomic spin pattern leads to an optical spin pattern which in turn sustains the atomic spin pattern by optical pumping. As the propagation time of light in the feedback loop is only of the order of a few nanoseconds, the feedback is essentially instantaneous compared to the microsecond to millisecond time scale of the atomic dynamics (this assumption holds fairly nicely even for the typical time scale of tens of nanoseconds for the electronic nonlinearity). Hence the dynamics of the light field can be adiabatically eliminated and the atomic degrees of free couple effectively to themselves via a light-mediated interaction. A similar argument is used for self-organization in transversely pumped cavities (e.g.~\cite{nagy10,baumann10,ritsch13}) and is of course at the heart of many schemes for quantum simulation of many-body systems. In order to investigate the connections to lattice-models for condensed-matter problems further and to clarify the local and global aspects of the diffractive coupling, we are going to look now into localized excitations and perturbations.


\begin{figure}[h]
\unitlength1mm
\begin{picture}(150,100)
\put(0,50){\includegraphics[width=7.5cm]{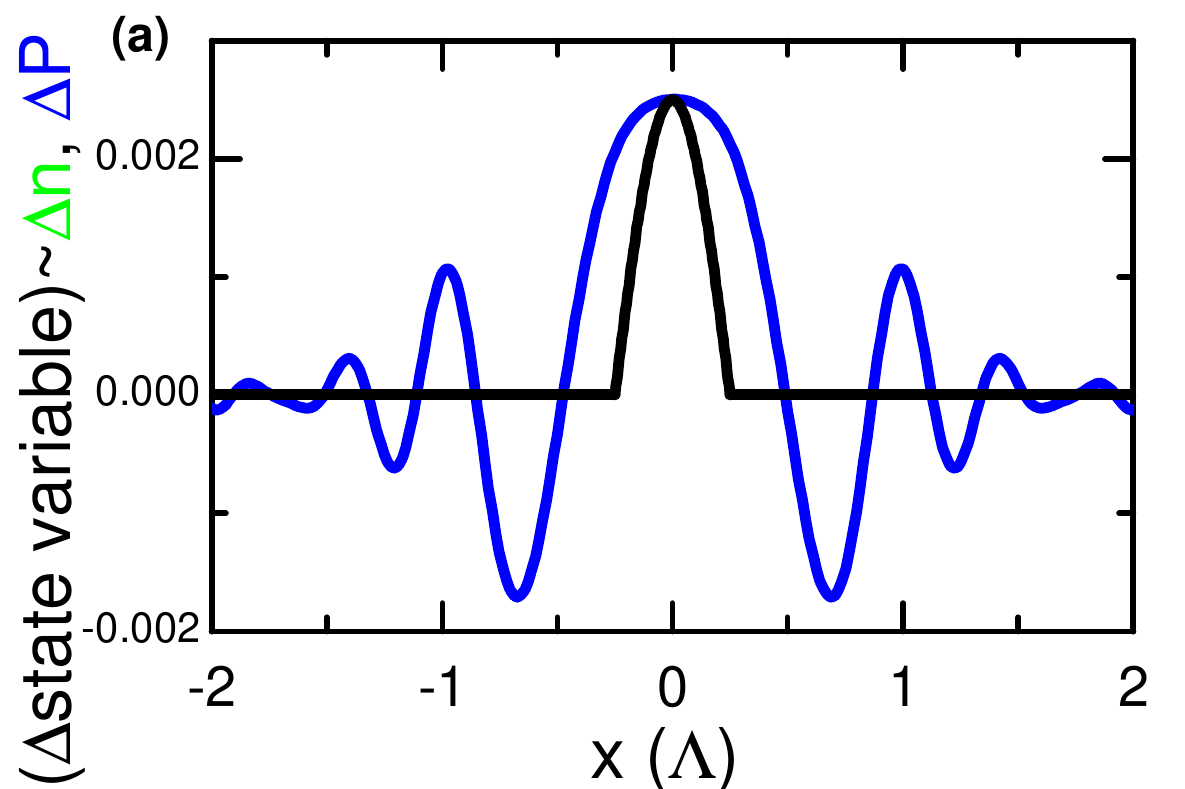}}
\put(75,50){\includegraphics[width=7.5cm]{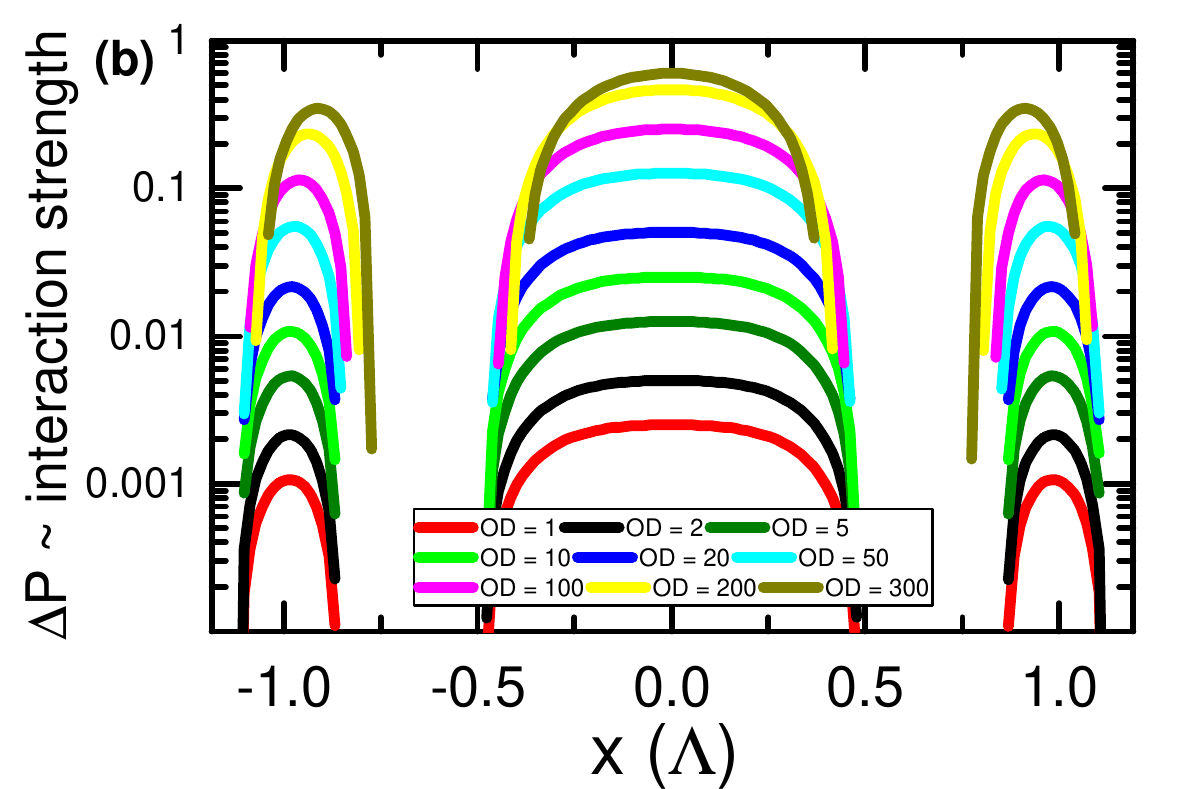}}
\put(0,0){\includegraphics[width=7.5cm]{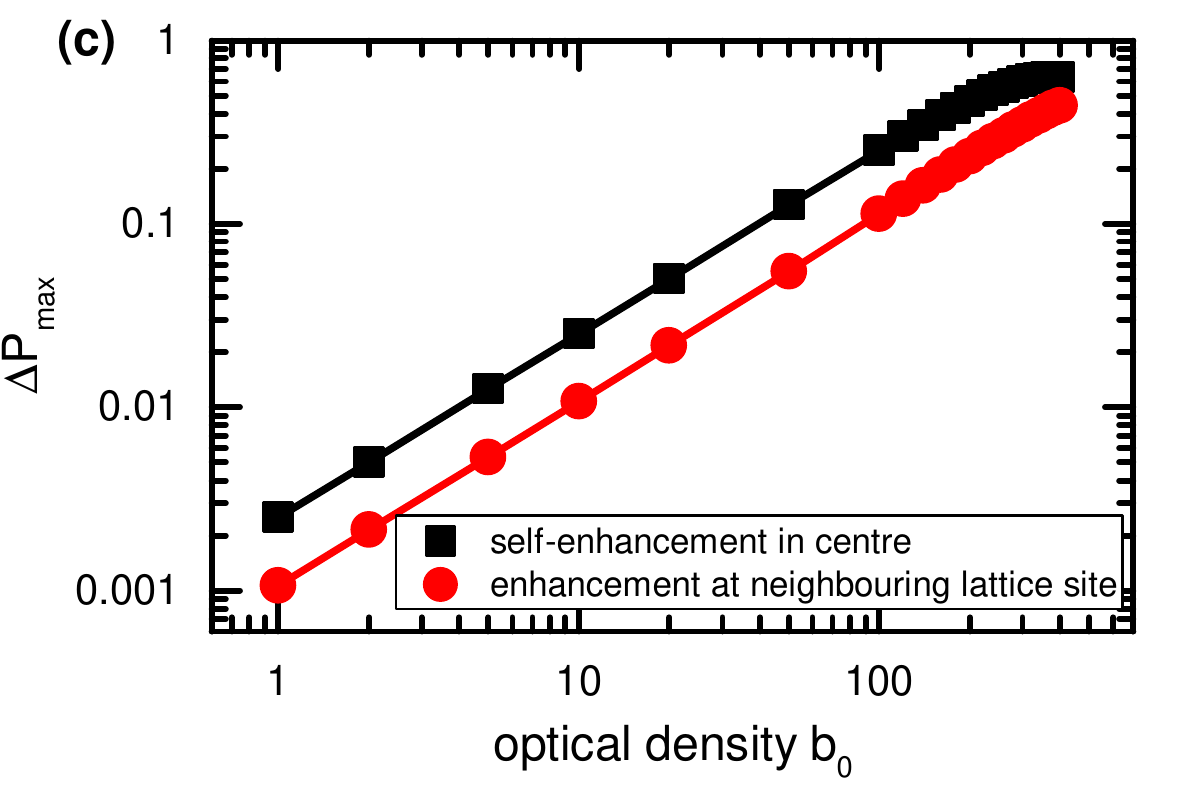}}
\put(75,0){\includegraphics[width=7.5cm]{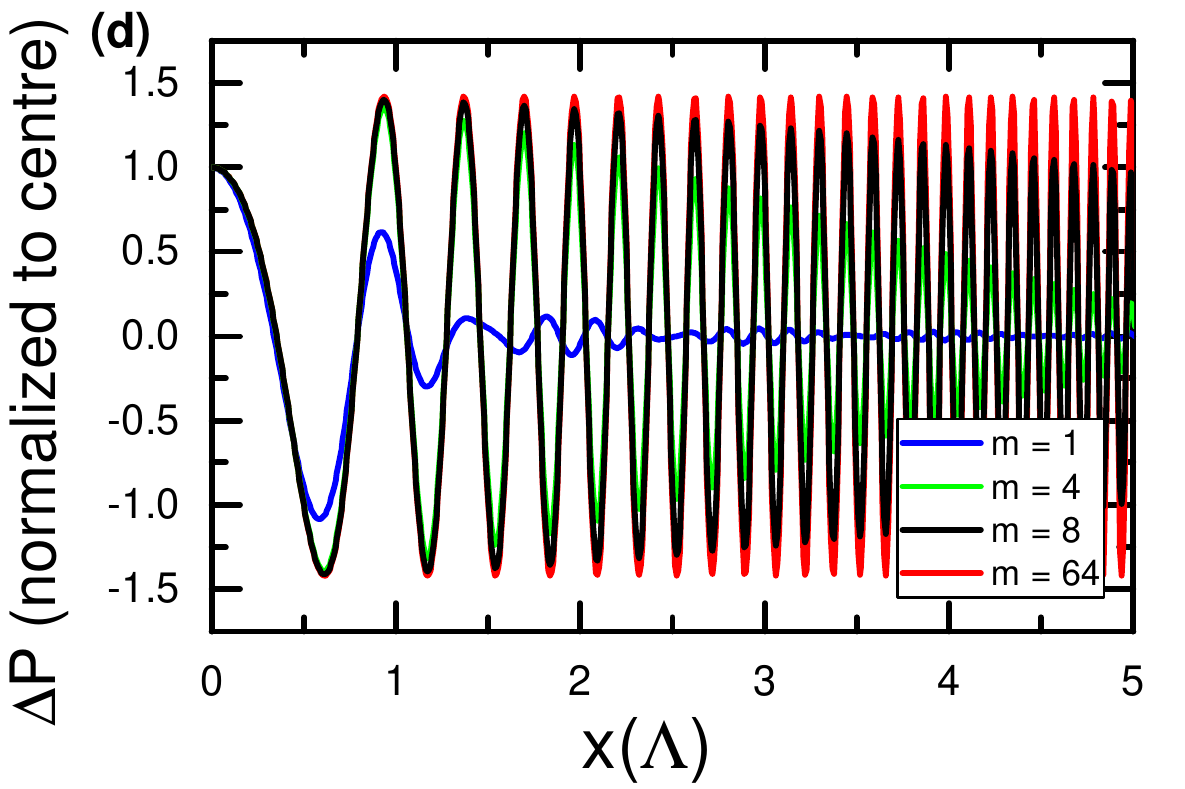}}
\end{picture}
\caption{\label{fig:kernel}\textbf{(a)} 1D cut through 2D profiles of localized perturbation of atomic state (black, scaled to match the feedback pump rate in peak for presentation purposes) and resulting perturbation (i.e.\ without the homogeneous background) in the  pump rate of backward beam (blue) after one round trip in feedback loop. The central lobe leads to the self-enhancement of the instability, the first side lobes to an enhancement at neighbouring lattice sites. Transverse space is measured in periods of the periodic lattice. The calculations are done in 2D and only profiles are presented. \textbf{(b)} Feedback pump rate for different optical densities in line centre $b_0$ in logarithmic scale. \textbf{(c)} Peak values of central (black squares) and first lopes (red circles) for different optical densities in line centre $b_0$ in logarithmic scale. For all panels the atomic perturbation has been small (0.1), homogeneous input pump rate $P=1$, detuning $7\Gamma$, $R=1$. $b_0=1$ in \textbf{(a)}. \textbf{(d)} 1D calculations of perturbation of backward pump rate being a factor of $m$ narrower than in \textbf{(a)}. Results are then normalized  by dividing by pump rate in centre $x=0$.}
\end{figure}

For this, the homogeneous state of the atomic variable is perturbed by a small localized perturbation with a width of half the lattice constant. For simplicity, we take half a period of a harmonic wave (black line in Fig.~\ref{fig:kernel}\textbf{(a)}). For the 2D case, a perturbation with circular symmetry is created by rotating the profile. For the optomechanical case and red detuning, the black line illustrates a peak in density and a peak in the resulting refractive index distribution (for positive detuning, the density perturbation would be a trough, not a peak, but the refractive index perturbation is the same). A homogeneous pump beam is impinging onto this perturbation. The blue line displays the resulting perturbation in the backward beam. It shows a central lobe surrounded by damped oscillations, as typical for diffractive ripples in Fresnel diffraction. The positive value at the centre leads to positive feedback at the site, i.e.\ to the self-enhancement of the instability. The first negative ripple will push atoms out of the vicinity of the central lattice site and deplete the area of atoms (for red detuning). In contrast, atoms will bunch again at the position of the first ripple maximum  (about) one lattice period apart. This will lead to the formation of a density modulated state with a local periodicity. Above threshold, it will lead to the emergence of a periodic density modulated state or atomic crystal, in the long-term dynamics.  For positive detuning, the negative ripple is attractive and will lead to bunching of atoms around the central optical peak, but expel atoms from lattice sites a period away. In 2D, this leads to the honeycomb structures in the atoms discussed in Sec.~\ref{fig:feedback_optomech}.

For a formed crystal, a localized (or wave-like extended) perturbation will spread across lattice sites via nearest neighbours. This creates the exciting possibility of stimulating density waves propagating across the structure. As the optical lattice is self-induced and hence dynamic, one can expect a richer dynamics of excitations than accessible in externally imposed, static optical lattices. At some finite distance above threshold even a band of unstable wavenumbers is present, which might make some aspects of longitudinal sounds accessible. It remains to be seen how strongly these degrees of freedom are damped.

For the magnetic dipole case, the black line in Fig.~\ref{fig:kernel}\textbf{(a)} represents a positive perturbations in the orientation $w$ and the blue line the resulting perturbation in the difference pump rate. The negative ripple is responsible for the anti-ferromagnetic coupling leading to negative orientation half a lattice period apart. The first positive ripple will support the first recurrence of the positively aligned state of the antiferromagnet. As argued in \cite{kresic18}, this has strong similarity to the dynamics of Ising spin on a lattice. `Superspins' on the self-induced lattice will couple to neighbouring sites similar to Ising spins with antiferromagnetic coupling. As the amplitude of the ripples decreases fast and their spatial frequency becomes incommensurate with the lattice period for distances larger than a lattice period, the coupling is essentially a nearest-neighbour coupling. We will come back to this question in a few paragraphs. After a local perturbation, one can expect the propagation of spin waves over the lattice. Note that this is a `simulation' of magnetic ordering by real spins and controlled by real magnetic fields, not pseudospins controlled by pseudo-magnetic fields. Only the coupling is mediated by light and enables magnetic ordering via light-mediated interactions in a density and temperature regime between solid-state systems (very dense, usually quite hot) and spinor condensates (very dilute, very low temperature).

Fig.~\ref{fig:kernel} \textbf{(b)} indicates that the interaction strength increases with optical density. This increase is linear in a wide range of optical densities (Fig.~\ref{fig:kernel}\textbf{(c)}). This is of course expected from Eqs.~(\ref{eq:chi_optomech}) and (\ref{eq:chi_mag}), (\ref{eq:chi_maglin}). From these equations it is clear that the coupling strength will be determined by the linear phase shift $\phi_{lin}$ or dispersive optical density, i.e.\ will be controlled by optical density and detuning. The discussion of thresholds and critical optical densities in the sections before also supports the notion that the dispersive optical density is the essential coupling parameter, although obviously there is no coupling without a pump.  As there is no feedback without reflectivity $R\phi_{lin}$ is a suitable cooperativity parameter for coupling in the single-mirror feedback experiment (see also the driving term in Eq.~(\ref{eq:eta_disp})). This matches optical density in line centre for collective scattering and superradiance \cite{bonifacio78a,guerin16} and optical density divided by cavity losses for cavity systems (e.g.\ \cite{bonifacio78a}).

As apparent from  a close inspection of Fig.~\ref{fig:kernel}\textbf{(a)} and of Fig.~\ref{fig:kernel}\textbf{(b)} the strongest positive diffractive ripple is displaced slightly from lattice period 1 (and the negative one from period 0.5).  This does not invalidate the argument for an Ising-like lattice model as on the formed lattice the signs of the coupling will be preserved. This displacement increases for increasing optical density (Fig.~\ref{fig:kernel} \textbf{(b)}). This is due to higher harmonics being important in the exponential obtained from integrating the paraxial wave equation (\ref{eq:chi_mag}). For the same reason the self-enhancement coefficient (amplitude of the central lobe) and the coupling coefficient to the next neighbour (amplitude of the first diffractive ripple) saturate and then decrease again with increasing optical density (Fig.~\ref{fig:kernel}\textbf{(c)}, for the amplitude of the first diffractive ripple saturation just sets in at the highest density displayed). At the same time, the relative amplitude of the oscillations further out increases. The behaviour is qualitatively the same in 1D and 2D but saturation sets in slightly earlier in 1D and amplitude and frequency of oscillations are slightly different.

Another way to increase the amplitude and spatial range of the coupling oscillations is to make the perturbation narrower. Fig.~\ref{fig:kernel}\textbf{(d)} shows the coupling induced by the perturbation of the backward beam if the size of the perturbation is reduced by a factor $m$ compared to Fig.~\ref{fig:kernel}\textbf{(a)} (the calculations are done in 1D to achieve better resolution). Making the perturbation more localized, the relative amplitude of oscillations to peak value of central lobe increases, the damping of oscillations becomes smaller, i.e.\ significant oscillations are observed at higher distances, and their frequencies increase. For a perturbation by a Delta-function, one expects that the perturbation of the  backward pump rate approaches the Fresnel diffraction kernel (e.g.\ \cite{siegman86}) proportional to $\cos{(2\pi r^2)}$. The same conclusion was reached before in \cite{zhang18} by assuming that the susceptibility is small enough to expand the exponential solution of Eq.~(\ref{eq:chi_optomech}) to first order, reinserting it into the equation of motion for the dynamics of the BEC and then integrating it out assuming $d \gg \lambda$. This means that the interaction is infinitely-ranged and highly oscillatory leading to the possibility of a wealth of potential structures. However, as the frequency of oscillations increases with distance, for an excitation of finite size the interaction potential averages out at large distances but interesting shorter range interactions, going beyond next neighbour, remain \cite{zhang18}. This matches the conclusion reached from the direct discussion of a localized, but finite size perturbation in Figs.~\ref{fig:kernel}\textbf{(a)}, \textbf{(b)}, which had been proposed for atomic superspins for the case of magnetic ordering before in \cite{kresic18}. For the quantum degenerate case, the resulting structures include chains and lattice of quantum droplets \cite{zhang18}. For a more general model complex superlattice structures have been predicted \cite{zhang20}. Their properties are controlled by BEC density, scattering parameter and strength of external driving.  As noted, also in the case with thermal atoms the range of interaction will depend on the localization of the central peak depending on parameters. The quasiperiodic and superlattices structures observed in hot vapours as discussed in Sec.~\ref{sec:sym} should be related to the same mechanism although their interpretation was done in terms of wave mixing phenomena and not interaction potentials.

\section{Self-organization via diffractive coupling in cavities}
\begin{figure}[h]
\includegraphics[width=15cm]{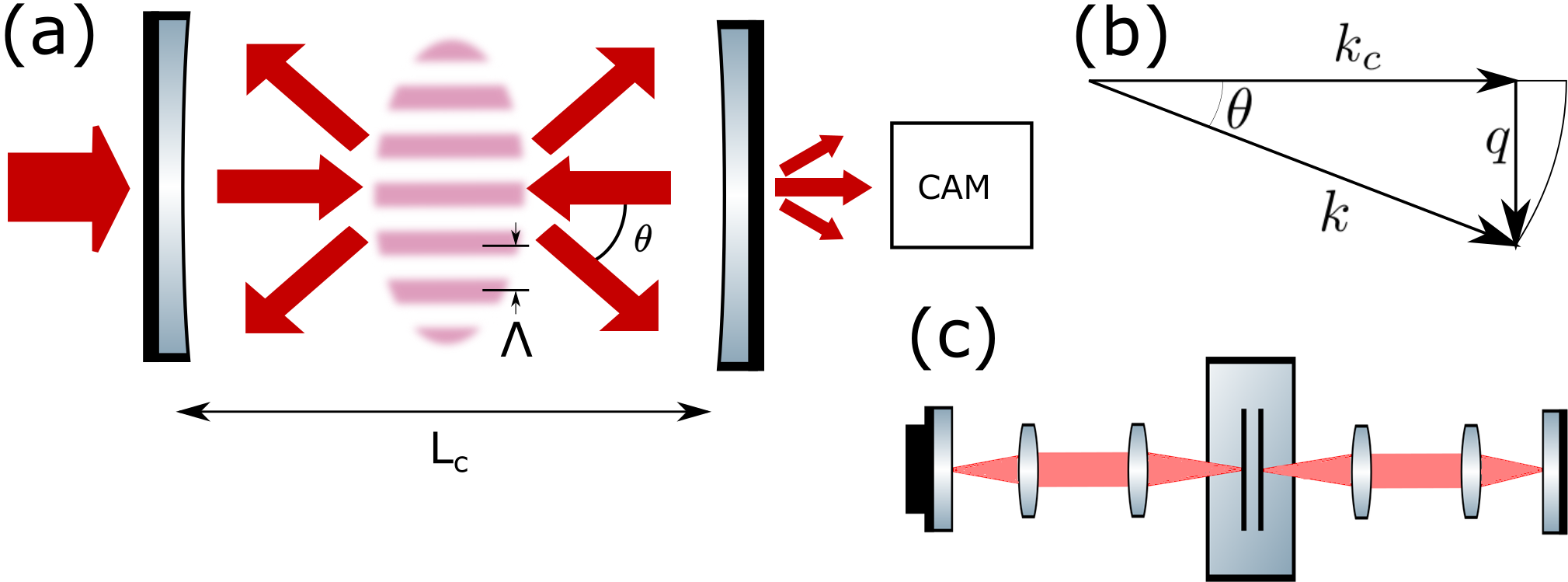}
\caption{\label{fig:cavityscheme}
\textbf{(a)} Scheme of a cavity of length $L$ driven by a coherent input field and containing a cloud of cold atoms in the centre. A modulation of an atomic state variable of period $\Lambda$ leads to scattering of the pump into sidebands at an angle $\theta$. The interference of these sidebands sustains the structure in the cloud.  \textbf{(b)} Light with a wavevector $k$ larger than the resonant cavity wavevector $k_c$ can reestablish resonance by tilting to a suitable angle $\theta$. \textbf{(c)} Plano-planar cavity with two afocal telescopes with diffractive properties equivalent to plano-planar cavity of reduced diffractive length.
 }
\end{figure}

Another important scheme for self-organization by diffractive coupling is a cavity filled with a nonlinear medium and driven along the longitudinal axis by a coherent input beam. This has been intensively investigated in optical pattern formation and as it is hard to give a comprehensive account of the literature. We mention only \cite{lugiato87,tlidi93,firth94c,martin96} as important milestones for the theoretical analysis. Fig.~\ref{fig:cavityscheme}\textbf{(a)} depicts the scheme. The atomic cloud is placed in a plano-planar resonator (or possibly a degenerate resonator with curved mirrors). It is driven by a coherent input beam off-resonant to the cavity. Although absorptive pattern formation is possible \cite{firth94c} the more attractive option is to use the dispersive limit of large detuning \cite{lugiato87}.  Any periodic spatial modulation in the atomic cloud will lead to the scattering of pump photons into spatial sidebands. The interference of these sidebands with each other will then sustain again the structure in the cloud. The sideband with the lowest threshold will be the one which is resonant to the cavity. Fig.~\ref{fig:cavityscheme}\textbf{(b)} illustrates how the angle of the sideband $\theta$ is determined by the condition that the projection of the wavevector $k$ on the cavity axis needs to match the resonant cavity wavevector $k_c$. If $\delta \phi$ denotes the phase mismatch between pump and cavity wavevector, this gives in second order of $\theta$ the condition
\begin{equation}\label{eq:phasemismatch}
  \delta \phi = (k-k_c) 2L_c = k2L(1-cos{\theta}) \approx kL \theta^2  = \frac{q^2}{k}\, L_c ,
\end{equation}
where use was made of (\ref{eq:angle}) to connect sideband angle and transverse wavenumber. This illustrates that the diffractive phasor in Eq.~(\ref{eq:parax_Fourier}) compensates for the phase mismatch to the cavity resonance. The period of the structure is then given by
\begin{equation}\label{eq:period_cavity}
  \Lambda = \sqrt{\frac{2\pi L_c \lambda}{\delta \phi}} .
\end{equation}
As for the single-mirror and the counterpropagating beam case, it is proportional to the square root of the product of wavelength and longitudinal system size but not depending on transverse boundary conditions.


In \cite{tesio12}, the possibility of self-organization of cold thermal atoms in a longitudinally pumped resonator is demonstrated for blue and red detuning. The patterns selected have a hexagonal symmetry. For red detuning, both density and light intensity show hexagonally coordinated peaks, whereas for blue detuning the structures are complementary, i.e.\ honeycombs for the density and hexagonally coordinated peaks for the light intensity, as one might expect from the discussion of the single-mirror feedback system. The quantitative analysis in \cite{tesio12,costa-boquete20} indicates that threshold should be achievable in a cavity of only a moderate Finesse of 30 compatible with a cavity placed outside the vacuum chamber, if the anti-reflection coatings of the cell windows are of high quality. In \cite{baio20} pumping with a vortex beam carrying orbital angular momentum is considered. This leads to concentring rings of density peaks rotating around the centre and creates interesting opportunities for transport of self-organized excitations.

Due to the high diffraction losses in planar-planar cavities of macroscopic length, experimental observations have been limited to nonlinear media as liquid crystals and semiconductor quantum wells allowing very short cavities \cite{kreuzer90,ackemann00a,hoogland02} and patterns are usually quite irregular due to spatial inhomogeneities. Many aspects of the diffractive coupling in plano-planar cavities can be recovered in mode-degenerate cavities with curved mirrors (as the confocal or concentric cavity) or cavities with intra-cavity lenses \cite{arnaud69}. Most work was done in lasers and photorefractive oscillators\cite{huyet95,staliunas97c}, but few-mode structures were obtained in a confocal resonator with hot sodium vapour \cite{lippi94a,lippi94b}. Multi-mode transversely pumped cavities use also degenerate situations like the confocal one \cite{kollar17}. A configuration like in Fig.~\ref{fig:cavityscheme}\textbf{(c)} borrows from the `virtual mirror' concept discussed for the single-mirror feedback system in Sec.~\ref{sec:principle}. Inserting  two afocal telescopes into a long plano-planar cavity yields a cavity with a much shorter effective diffractive length given by the distance of the focal points of the two inner lenses. This allows to control length scale and limit diffraction losses while keeping the modal properties of the plano-planar resonator. The calculations in \cite{costa-boquete20} indicate that the relatively high losses introduced by the residual Fresnel reflection losses from the surface of four anti-reflected coated lenses might be just tolerable. Otherwise, one can use concentric or confocal configurations without intracavity lenses.

\begin{figure}[h]
\centering
\includegraphics[width=9cm]{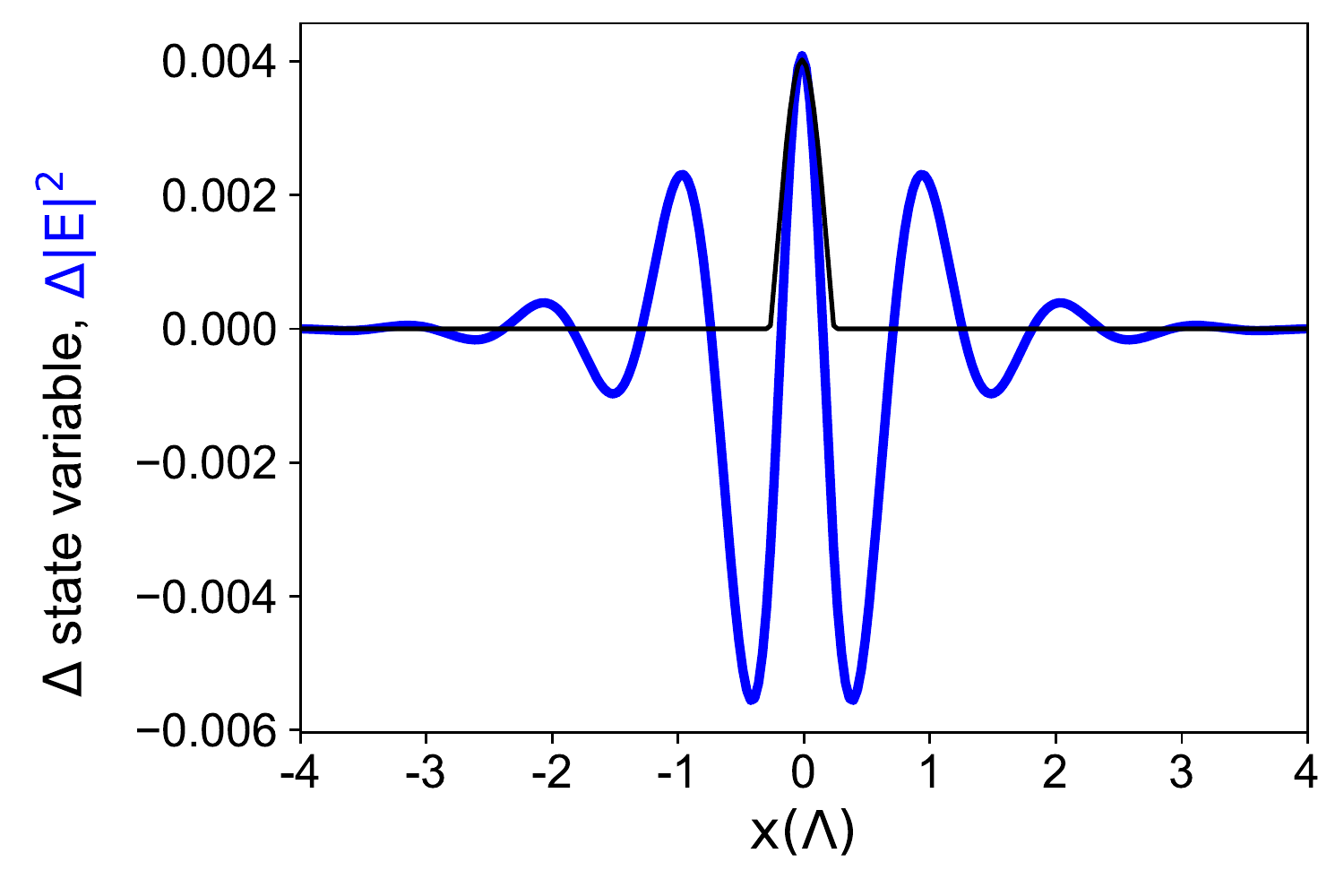}
\caption{\label{fig:cavity_coupling} Perturbation of atomic state variable (black line) proportional to refractive index and resulting perturbation (scaled to meet perturbation amplitude for purpose of visualization) in intra-cavity intensity after a time corresponding to five times the inverse of the cavity decay rate (see text and Fig.~\ref{fig:kernel}) for details of the procedure). Parameters: pump intensity 1.05 above threshold; cavity detuning zero, $\Delta = -10 \Gamma $, cooperativity parameter $C = 0.0125$. }
\end{figure}

The structures investigated in \cite{tesio12} are coupled light-matter structures. Although most of the work in optical pattern formation concentrated on optical wave mixing aspects and very often the material dynamics was adiabatically eliminated \cite{lugiato87,tlidi93,firth94c,martin96}, one can make an argument for light-mediated interaction between atoms also for longitudinally pumped cavity schemes, in particular for ones with relatively low Finesse. Following the approach discussed in Sec.~\ref{sec:coupling}, Fig.~\ref{fig:cavity_coupling} shows a calculation in which the density of the atomic cloud (black line) is slightly perturbed from the homogeneous state by a localized peak. Then the field equation is integrated (keeping the density constant) for a time corresponding to the photon lifetime in the cavity. The resulting perturbation (blue line) has a central lobe (providing self-enhancement) and is surrounded by damped oscillations. The first diffractive ripple peaks nearly at the lattice period. The higher order ripples are slightly further off the lattice site than the first one but closer aligned with the lattice than in the single-mirror system. These considerations indicate the possibility of a lattice-type interaction on the self-induced lattice also for cavity systems. Further details are currently under investigation.

\section{Conclusions and outlook}
In this contribution, we analyzed a conceptually very simple system for atomic interaction mediated by light, which consists of a laser driven ensemble of atoms and a single, plane retro-reflecting mirror. The length scale of the transverse instability is given by diffractive dephasing and the Talbot effect and can be varied by adjusting the mirror distance. Light-atom coupling by external and various internal degrees of freedom can be used to induce light-mediated atomic interactions. For the most interesting case of dispersive interactions in the large detuning limit, the linear phase shift of the cloud times reflectivity of feedback mirror is a suitable cooperativity parameter. In the limit $R\to 1$, the cooperativity parameter is given by the linear phase shift, i.e.\ the dispersive optical density, alone. Utilizing the archetypical 2-level electronic nonlinearity implies always a finite intensity threshold and demands relatively high optical densities as the strength of driving and relaxation depends on the dipole matrix element in the same way. This is different for the coupling relying on optical pumping and optomechanical interactions, which can in principle approach zero intensity threshold (or the lowest threshold compatible with quantum effects, see below), if the temperature is low enough and other parasitic relaxation channels can be controlled. The requirements on optical density are strongly reduced, but the requirement of a finite optical density remains.

Coupling of the light degrees of freedom on the Poincar{\'e} sphere with the magnetic moments in the atomic ground state via optical pumping nonlinearities leads to spontaneous magnetic ordering. Light helicity ($S_3$ Stokes parameter) couples to the atomic dipole, linear polarization ($S_1$, $S_2$ Stokes parameters) to an atomic quadrupole represented by a $\Delta m=2$ coherence, where the relative phase of the $\sigma$-light components relates to polarization direction on the one hand and to quadrupole direction, again given by the phase of the coherence, on the other hand. We presented evidence of a self-organized structure in a quantum coherence. Interaction between structured light and structured coherences is interesting for storing images and for metrology in quantum technologies.

As there is renewed interest in hot vapours (e.g.\ \cite{whiting17,santic18,fontaine18}), we comment that all instabilities based on internal degrees of freedom were already demonstrated to exist or should be accessible in suitable prepared situations. Achieving a high optical density is actually easier in hot vapours but Doppler broadening, ballistic transport, and potentially the presence of many hyperfine lines reduce effective interaction and make modelling much more demanding. Introducing a buffer gas can help to maintain homogeneous broadening, potentially mask hyperfine splitting and lead to diffusive motion which can be modelled easily. However, the strength of the light-matter interaction will decrease due to the collisional broadening.  In contrast, cold atoms have negligible Doppler broadening and the maximum light-matter coupling allowed by the natural lifetime.

The decisive novel feature in cold atoms is the coupling to external degrees of freedom of the atoms, i.e.\ the centre of mass motion, via the dipole force creating an optomechanical nonlinearity. This leads to bunching of atoms to density modulated structures. Current experiments are in qualitative agreement with a description based on a conservative Vlasov equation. The lifetime of the structures in the experiment is probably limited by heating effects and the radiation pressure force. It will be interesting to study the influence of these effects theoretically and to look at the possibly of cooling in a red detuned situation, in analogy to cavity cooling \cite{domokos02,black03,jaeger17,jungkind19}. Introducing controlled velocity damping via external molasses beams is interesting to study the influence of the gradual introduction of dissipation into self-organization and to reach finally the soft-matter case at high dissipation. The latter case was recently implemented using dielectric beads in a water solution \cite{bobkova21}.

We commented on the connections to other schemes, longitudinally pumped cavities and counterpropagating beams, in which self-organization occurs due to diffractive coupling. All these system have in common that their critical transverse length scale is of the order of the square root of the optical wavelength times the longitudinal system size, a property simply arising from the scaling properties of the paraxial wave equation (\ref{eq:parax_real}). In the transverse directions there are no boundary conditions, in principle, and two continuous symmetries -- the rotational and the translational symmetry in the transverse plane -- are broken simultaneously and spontaneously, as far as possible within the limitation of the experimental realization. On the fundamental side, this creates exciting opportunities to study the Kibble-Zurek mechanism in 2D for different kind of structures. It also opens up a path to 2D supersolids (see below). From the photonics and quantum optics point of view, a long-term vision is to achieve all-optical manipulation at single-photon level by coupling to the Goldstone modes arising from spontaneous symmetry breaking \cite{greenberg11,dawes05,gupta07}.

Moving over to quantum degenerate gases as the media of interest creates further exciting possibilities as the structured state will break translational and phase symmetry at the same time and hence provides a path towards a 2D supersolid in addition to dipolar quantum gases \cite{zhang19,zhang21,hertkorn21,norcia21}. Numerical simulations starting from plane wave initial conditions for a condensate wavefunction coupled to the optical feedback but without contact interactions yield a structured state indicating the possibility of a 2D dynamic supersolid. By projection on fundamental momentum modes a mapping to the Dicke phase transition is possible in a similar way to the situation of transversely pumped cavities. Other investigations \cite{zhang18,zhang20} indicate that for a BEC with repulsive interaction and a finite number of atoms a single droplet or chains and 2D partitions of droplets are the ground state of the dynamics, partially displaying very complex and beautiful superlattice structures. The connection between these two approaches needs to be explored in future work.

The diffractive interaction leads to an oscillatory interaction potential. For a single atom constituting a $\delta$-function in space, this interaction has infinite range, but the period of oscillations is rapidly increasing with distance. As a result, the interaction of a cluster or droplet of atoms centred of finite width with the surrounding is effectively short range, with the details depending on the width and the profile of the distribution. Nevertheless, the complex shape of the remaining oscillations at finite range leads to the unusual symmetries and superlattices investigated in \cite{zhang18,zhang20} and is probably related to the quasiperiodic and superlattice structures experimentally and theoretically observed in hot vapours before. Atomic clusters of roughly the size of half the lattice period show effectively a next neighbour interaction on the self-induced lattice. For the case of magnetic ordering, these clusters can be regarded as a kind of `superspins' interacting on the self-induced lattice. This provides a strong analogy to lattice Ising models.

One further interesting area of future investigation is the relation between instabilities based on internal and external degrees of freedom, in particular the interaction of magnetic and density ordering in thermal atoms and feedback-induced spinor dynamics and momentum dynamics in quantum degenerate gases, e.g.\ \cite{mivehvara21,masalaeva21}. As a starting point for the former,  Ref.~\cite{schmittberger16} analyzes the case of counterpropagating beams for a $J=1/2 \to J'=3/2$- transition and points out that optical pumping and polarization dependent optical potentials can provide simultaneously magnetic ordering and atomic bunching in the longitudinal grating. By the same argument, one can expect a mutual enhancement also in the transverse case, i.e.\ the magnetic superspins should become spikier than from the optical pumping alone due to the optomechanical bunching.

We conclude with an outlook on the connection between the cases of transverse and longitudinal self-organization. We consistently found that the structures observed can be qualitatively understood by taking only the transverse gratings (`transmission gratings' in the language of grating theory) into account. This does not exclude that longitudinal, wavelength-scale gratings are of quantitative importance as it has been argued for optomechanical structures in counterpropagating beams \cite{greenberg11,schmittberger16} and for electronic structures for the single-mirror feedback system \cite{firth17}.
However, these longitudinal structures are already induced by the counterpropagating  pumps. To obtain a spontaneous longitudinal instability as the CARL instability pioneered by Rodolfo Bonifacio coupled to a transverse instability, one would need to consider an extension of the multi-mode ring cavity model in \cite{tesio12} to include the possibility of spontaneous back scattering. Recent progress on multi-mode CARL instabilities in free space is documented in this special issue \cite{gisbert20}.

\vspace{6pt}

\acknowledgments{We are grateful to Enrico Tesio, Pedro Gomes, Abdoulaye Camara, Aidan Arnold and Alison M. Yao for fruitful discussion and collaboration.

\textbf{Funding Information:} This research was performed in the framework of the European Training Network ColOpt, which is funded by the European Union (EU) Horizon 2020 programme under the Marie Skłodowska-Curie action, grant agreement 721465.
The APC was funded by this grant. The collaboration between the two groups is further supported by the CNRS via the  Laboratoire international associé (LIA) `Solace', and the Global Engagement Fund of the  University of Strathclyde. Josh G. M. Walker is supported by a John Anderson Research Award from the University of Strathclyde, Ivor~Kre\v{s}i\'{c} has been supported by a University Studentship of the University of Strathclyde and the Leverhulme Trust. Earlier stages of the investigations at Strathclyde were funded by the Leverhulme Trust and  the Royal Society (London).

\textbf{Conflicts of interest:} The authors declare no conflict of interest.}

\end{document}